\documentclass[aps,prb,twocolumn]{revtex4}
\usepackage{amsfonts}
\usepackage{amscd}
\usepackage{bm}
\usepackage{amsmath}
\usepackage{amssymb}
\usepackage{color}
\usepackage{mathptmx}
\def\nicefrac#1#2{\frac{#1}{#2}}
\usepackage{ifpdf}
\usepackage[pdftex]{graphicx}
\DeclareGraphicsExtensions{.eps, .pdf, .jpg, .tif}
\usepackage{hyperref}
\usepackage{bookmath}
\usepackage{mathfrak}
\usepackage{mymatrices}
\def\cHb{\whH}
\newcommand{\whGR}{\widehat{\mathfrak{G}}^{\text{R}}}

\newcommand{\bpm}{\begin{pmatrix}}
\newcommand{\epm}{\end{pmatrix}}
\newcommand{\abs}[1]{\left|#1\right|}
\def\orbital{{\tt SO(3)_{\text{L}}}}
\def\spin{{\tt SO(3)_{\text{S}}}}
\def\gauge{{\tt U(1)_{\text{N}}}} 
\def\parity{{\tt P}} 
\def\time{{\tt T}}
\def\charge{{\tt C}} 
\def\conjugate{{\tt K}}
\def\spinorbit{{\tt SO(3)_{\text{L}+\text{S}}}}
\def\twoD{{\tt SO(2)_{\text{L}_{\text{z}}+\text{S}_{\text{z}}}}}
\def\piz{{\tt U}_{\text{z}}(\pi)}
\makeatletter
\newcommand{\rom}[1]{\expandafter\@slowromancap\romannumeral #1@}
\makeatother
\begin{document}
\title{
\vspace*{-20mm}
\hspace*{-10cm}{\small Published in Phys. Rev. B 88, 184506 (2013) 
[\href{http://dx.doi.org/10.1103/PhysRevB.88.184506}{doi: 10.1103/PhysRevB.88.184506}].
}\\
\vspace*{5mm}
Majorana Excitations, Spin- and Mass Currents on the Surface of Topological Superfluid $^3$He-B}
\author{Hao Wu} 
\author{J.A.Sauls}
\affiliation{Departments of Physics and Astronomy, Northwestern University, Evanston, IL 60208}
\date{\today}
% \pacs{67.30.hb, 67.30.hr, 67.30.ht}
%-----------------------------
\begin{abstract}
The B-phase of superfluid \He\ is a 3D time-reversal invariant (TRI) topological superfluid with an isotropic
energy gap, $\Delta$, separating the ground-state and bulk continuum states. 
We report calculations of surface spectrum, spin- and mass current densities originating from the Andreev surface
states for confined \Heb.
The surface states are Majorana Fermions with their spins polarized transverse to their direction of propagation
along the surface, $\vp_{\parallel}$. 
The negative energy states give rise to a ground-state helical spin
current confined on the surface.
The spectral functions reveal the subtle role of the spin-polarized surface states in relation to the ground-state
spin current.
By contrast, these states do not contribute to the $T=0$ mass current. 
Superfluid flow through a channel of confined \Heb\ is characterized by the
flow field, $\vp_s=\frac{\hbar}{2}\grad\varphi$.
The flow field breaks $\twoD$ rotational symmetry and time reversal ($\time$). 
However, the Bogoliubov-Nambu Hamiltonian remains invariant under the combined symmetry, $\piz\times\time$, where
$\piz$ is a $\pi$ rotation about the surface normal.
As a result the B-phase in the presence of a superflow remains a
topological phase with a gapless spectrum of Majorana modes on the surface.
Thermal excitation of the Doppler shifted Majorana branches leads to a power law suppression of the superfluid mass
current for $0 < T\lesssim 0.5 T_c$, providing a direct signature of the Majorana branches of surface excitations
in the fully gapped 3D topological superfluid, \Heb.
Results are reported for the superfluid fraction (mass current) and helical spin current for confined \Heb, including the
temperature dependences, as well as dependences on confinement, pressure and interactions between quasiparticles.
\end{abstract}
%-----------------------------
\maketitle
\section{Introduction}

A universal feature of superconductors and superfluids in which the ground state breaks one or more space-time symmetries,
in addition to $\gauge$ gauge symmetry, is pair-breaking at non-magnetic boundaries.
Pair-breaking results in Fermionic states that have energies within the continuum gap and are confined near impurities,
interfaces and topological defects such as vortices or domain walls.\cite{bal06,sau09,sal88}
Superfluid \He\ is the paradigm for BCS pairing with complex symmetry breaking.\cite{vollhardt90}
The B-phase of superfluid \He\ is also a paradigm for time-reversal invariant (TRI) topological
order.\cite{qi09,vol09,qi11}
Indeed the symmetry exhibited by the ground-state of superfluid \Heb\ and
the non-trivial topological invariant of the Bogoliubov-Nambu Hamiltonian are intimately related.\cite{vol09,vol09a,ryu10}
As a consequence Fermionic excitations, surface Andreev bound states (ABS), extending to zero energy
with a Dirac-type spectrum are confined on the boundary of superfluid \Heb.

Although the spectrum of surface ABS in \Heb\ has been known theoretically for some time,\cite{buc81} these
predictions did not attract much interest until the discovery that surface ABS in high-temperature superconductors provided
a novel spectroscopy of the unconventional pairing symmetries.\cite{hu94,buc95a,mat95,cov97,fog97}
The first experimental evidence for surface ABS in \Heb\ was provided by transverse acoustic impedance measurements at
frequencies below the continuum edge for quasiparticle pair production, $\hbar\omega < 2\Delta$ by Aoki et al.\cite{aok05},
and measurements of their heat capacity by Choi et al.\cite{cho06}
These studies involved \Heb\ confined by a disordered quartz or metallic surface for which the surface Fermionic spectrum
was predicted theoretically to be a finite density of states filling the gap from the Fermi level up to a an energy scale
$\Delta'\lesssim\Delta$.\cite{nag98,vor03} The experimental studies confirmed a sub-gap Fermionic spectrum with a finite
density of states at the Fermi level, but they provided no information on the Dirac spectrum predicted for specularly
reflecting surfaces.
Recent impedance measurements carried out on transducers pre-plated with a thin layer superfluid \Hefour\ show a frequency
response that is interpreted as the evolution toward a Dirac spectrum of surface ABS in the presence of reduced
surface disorder.\cite{mur09}

That the surface Andreev states of \Heb\ are Majorana fermions, and are topological in origin, was recognized by
Volovik\cite{volovik03,vol09} and by Qi et al.\cite{qi09} This lead to a number of new theoretical analyses to identify the
Majorana spectrum on the surface of \Heb, as well as new experiments and proposals to detect these novel excitations and
understand the coherence and topological protection of emergent Majorana Fermions in superfluid
\Heb.\cite{chu09,nag09,miz12a,miz12}
The Majorana Fermions are spin polarized \emph{transverse} to their direction of propagation
along the surface, $\vp_{\parallel}$, with a linear dispersion relation, $\varepsilon_{\pm}(\vp) =
\pm\,c\,|\vp_{\parallel}|$.
The negative energy states, which are fully occupied in the ground state, generate a helical spin current confined on the
surface. Similar ground-state helical spin currents are a key signature of 3D TRI topological insulators, and have been
detected with spin-polarized angle-resolved photo-emission on Bi$_{1–x}$Sb$_x$.\cite{hsi09} The spin-current on the surface
of \Heb\ was first discussed by Zhang et al.,\cite{zha87} but was not connected to the Fermion spectrum, and 
so far has not been detected.

In this paper we discuss effects of surface scattering and quasiparticle interactions on the Fermionic spectrum, and
ground-state currents in the vicinity of boundaries confining the 3D TRI superfluid B-phase. The spin-current spectral
function reveals the subtle role of the ABS and continuum spectrum in determining the ground-state spin current.
We also discuss mass transport and the response of the surface spectrum in confined superfluid \Heb. Superfluid mass flow
occurs in response to a phase gradient, or flow field, $\vp_{s}=\frac{\hbar}{2}\grad\varphi$, where $\varphi(\vr)$ is the 
phase of the B-phase Cooper pairs.
The flow field breaks the $\twoD$ orbital rotation symmetry about the normal axis of confined \Heb, as well as time reversal
($\time$) and particle-hole ($\charge$) symmetry.
However, the Bogoliubov-Nambu Hamiltonian remains invariant under the combined symmetry, $\piz\times\time$, where
$\piz$ is a $\pi$ rotation about the surface normal.
As a result the B-phase in the presence of superflow remains a topological phase with a gapless spectrum of Majorana modes on
the surface.
And in contrast to the spin current, the negative energy bound states do not contribute to the ground-state mass current.
However, thermal excitation of the Doppler shifted Majorana branches leads to a power law suppression of the superfluid mass
current for $0 < T\lesssim 0.5 T_c$, providing a direct signature of the Majorana branches of surface excitations in the
fully gapped 3D topological superfluid, \Heb.
Quantitative results are reported for the mass current and helical spin current for confined \Heb, including the temperature
and pressure dependences, as well as dependences on confinement, and interactions between the Fermionic quasiparticles.
The results reported here are discussed in context with the recent results on helical spin currents 
by Tsutusmi et al.\cite{tsu12b}

Starting from the Bogoliubov-Nambu Hamiltonian in Sec. \ref{sec-Hamiltonian} we review the
symmetries and the topological winding number governing the surface spectrum of \Heb. 
In Sec. \ref{sec-Eilenberger} we introduce Eilenberger's quasiclassical equation for the Nambu propagator that is the basis
for investigating the spectrum of surface states, spin- and mass currents of confined superfluid \Heb. 
The Nambu propagator for quasiparticles and pairs is obtained for a specular boundary, and the surface ABS spectrum is
discussed in Sec. \ref{sec-DOS}.
In Sec. \ref{sec-Gap} we connect the surface ABS to pair-breaking of the Cooper pair spectral density.
In Sec. \ref{sec-Projection} we show explicitly how the Majorana property of the surface bound state spectrum is encoded in
the Nambu propagator by introducing Shelankov's projection operators to construct the Majorana Fermion spinors.
In Secs. \ref{sec-Spin_Current_DOS}-\ref{sec-Spin_Current_Ground-State} we discuss the spin- and spin-current spectral
densities, and the contributions to the ground-state spin-current from the surface ABS and continuum spectrum.
Results for the spin current at finite temperature are presented in Sec. \ref{sec-Spin_Current_vs_T}.
In Sec. \ref{sec-Mass_Current} we consider the effects of an imposed phase gradient across a channel of confined \Heb. The
breaking of time-reversal symmetry by the flow field, and its effect on the topological class and Majorana spectrum are
discussed in Sec. \ref{sec-Mass_Current_Symmetries}, while the mass current spectral function in the presence of superflow.
including the effects of quasiparticle molecular field interactions are described in Secs.
\ref{sec-Mass_Current_DOS}-\ref{sec-Fermi-Liquid_Effects}.
We begin with an introduction to broken symmetries in the B-phase of superfluid \He, its order parameter and residual
symmetry.

\section{3D Topological Superfluid \Heb}\label{sec-Hamiltonian}

The normal phase of \He\ is separately invariant under spin- and space (orbital) rotations, gauge symmetry, as well as parity and time-reversal. Thus, the maximal symmetry group of the normal phase
is\footnote{The weak breaking of $\orbital\times\spin$ rotation symmetry by the nuclear dipolar interaction is of order
$10^{-7}E_f$ and not relevant in terms of the symmetry and topological classification of \Heb\ considered here. See however, 
Ref. \onlinecite{miz12a} on the role of the dipolar energy in topological protection for \Heb\ in a Zeeman field. 
Particle-hole symmetry is broken at the level of $10^{-3}\Delta$, and this weak violation is similarly omitted here.}
\be
\tt G=\orbital\times\spin\times\gauge\times\parity\times\time
\,.
\ee

The superfluid phases of \He\ are condensates of p-wave ($\tt L=1$), spin-triplet ($\tt S=1$) Cooper
pairs,\cite{vollhardt90} described by a pairing gap matrix (order parameter) in
spin-space of the form,
\be\label{eq-gap_matrix}
\hDelta(\vp) 
= i\vec\vsigma\vsigma_y\cdot\vec{d}(\vp)
            = 
              \begin{pmatrix} - d_x + i d_y		&		d_z	\cr
								d_z					&  d_x + i d_y
			  \end{pmatrix}
\,,
\ee
where $\vec{d}(\vp)$ transforms as a vector under rotations in spin-space, and is a p-wave function of the
orbital momentum of the Cooper pairs. For inhomogeneous phases, e.g. flow states and  confined \He, the order parameter is
also a function of the center of mass coordinate, $\vR$, of the Cooper pairs.

The B-phase of superfluid \He\ is the realization of the Balian-Werthamer (BW) phase for a condensate of p-wave,
spin-triplet Cooper pairs with total angular momentum $\tt J=0$.\cite{bal63}
This phase is defined by $\vec{d} = \Delta\,\vp/p_f$, where the gap magnitude, $\Delta$, is real. The corresponding
pairing gap matrix, $\hDelta(\vp)=\Delta\,i\vec\vsigma\vsigma_y\cdot\vp/p_f$, is manifestly invariant under
\emph{joint} spin- and orbital rotations ($\tt J=0$).
Thus, the B-phase of \He\ breaks spin and orbital rotation symmetries, as well as parity and gauge symmetry, but
preserves time-reversal, particle-hole symmetry and \emph{joint} rotations of both the spin and orbital
coordinates, i.e. the residual symmetry group is
\be
\tt H = \spinorbit\times\time
\,.
\ee
The broken \emph{relative} spin-orbit rotation symmetry leads to a spontaneously generated spin-orbit coupling in
the B-phase with an energy scale of order $\Delta$, which plays a central role in determining the excitation
spectrum of superfluid \Heb.
The stiffness associated with relative spin-orbit rotations is responsible for \emph{transverse sound}, acoustic 
circular birefringence, and thus Faraday rotation of the polarization of transverse sound.\cite{moo93,lee99,sau00a}
Similarly, the Fermionic spectrum is defined by Bogoliubov quasiparticles which are \emph{momentum} and
\emph{helicity} eigenstates derived from the broken relative spin-orbit rotation symmetry.

The Fermionic spectrum of superfluid \Heb\ is governed by the $4\times 4$ Bogoliubov-Nambu Hamiltonian\footnote{We
use wide hats to denote $4\times4$ Nambu matrices and small hats to denote $2\times2$ spin matrices. We omit the
hats on the particle-hole ($\vec\tau$) and spin ($\vec\sigma$) Pauli matrices.}
\be\label{eq-Hamiltonian_B-phase} 
\hspace*{-4mm}
\cHb = \xi(\vp)\tauz + \whDelta(\vp)\,,\quad
\whDelta(\vp) = \begin{pmatrix}			0					& 	\hDelta(\vp)	\cr
									\hDelta^{\dagger}(\vp)	&		0
				\end{pmatrix}
\,,
\ee
where $\vec\tau = \{\taux,\tauy,\tauz\}$ [$\vec\vsigma = \{\vsigma_x,\vsigma_y,\vsigma_z\}$] are Pauli matrices in
particle-hole (spin) space, and $\xi(\vp)\equiv\vp^2/2m^{*}-\mu$ is the excitation energy of normal-state
particles and holes, and $\mu=\nicefrac{1}{2}\,v_f\,p_f$ is the Fermi energy at $T=0$, with $p_f$ and $v_f$ being
the Fermi momentum and Fermi velocity, respectively.
A unitary transformation of the Bogoliubov-Nambu Hamiltonian, $\whS\cHb\whS^{\dag}\rightarrow\cHb$, reduces Eq.
\ref{eq-Hamiltonian_B-phase} to the Dirac form,\footnote{The transformation is a $\pi$-rotation of hole spinors
relative to particle spinors: $\whS = \tinyonehalf(1 + \tauz) + \tinyonehalf(1 - \tauz)(i\vsigma_y)$.}
\be
\cHb = \xi(\vp)\tauz + c\,\vp\cdot\vec\vsigma\,\taux
\,,
\ee
in which the spin-orbit coupling of the ``relativistic'' Fermionic states is explicit, and the ``light'' speed,
which is determined by the bulk gap and Fermi momentum, $c = \Delta/p_f$, is three orders of magnitude below the
Fermi velocity.

The bulk excitation spectrum is obtained from $\cHb^2 = E(\vp)^2 = \xi(\vp)^2 + \Delta^2$. The negative energy
states are filled and account for the condensation energy of the B-phase ground state. Fermionic excitations in the
bulk phase are doubly degenerate helicity eigenstates with excitation energy, $E(\vp)=\sqrt{\xi(\vp)^2+\Delta^2}$,
that is gapped over the entire Fermi surface.

The B-phase belongs to \emph{topological} class DIII of 3D topological insulators and superconductors.\cite{sch08}
The topological invariant is the 3D winding number,\cite{sch08,vol09,vol09a}
\ber\label{eq-winding_number-3D}
N_{\text{3D}} &=& \int\frac{d^3p}{24\pi^2}\,\epsilon_{ijk}\,\mbox{Tr}\Big\{\Gamma\,(\cHb^{-1}\partial_{p_i}\cHb)
\\
&\times&(\cHb^{-1}\partial_{p_j}\cHb)(\cHb^{-1}\partial_{p_k}\cHb)\Big\}\,=\, 2
\nonumber
\,,
\eer
derived from, $\Gamma\cHb\Gamma^{\dag} = -\cHb$ with $\Gamma\equiv\charge\times\time$, where $\charge=\conjugate\times\taux$
is the conjugation symmetry of the Bogoliubov-Nambu Hamltonian.
At an interface between a topological quantum phase - with non-zero winding number and a bulk gap - and a 
non-topological phase the gap must close.\cite{sal88}
For \Heb\ Majorana Fermions with a linear dispersion relation, $\varepsilon^{\text{B}}_{\pm}({\vp}) = \pm
c\,|\vp_{\parallel}|$, form a ``Dirac cone'' of states at the interface, where $\vp_{\parallel}$ is the
momentum in the plane of the interface.

Consider \He\ confined in one dimension, i.e. a channel of width $D$. For channel widths $D\ge D_c\approx 9\xi_0$ the ground
state is a ``distorted'' B-phase defined by \cite{vor03}
\be\label{eq-d_vector_B-phase}
\vec{d} = \Delta_{\parallel}\left(\hp_x\,\hat\vx + \hp_y\,\hat\vy\right) + \Delta_{\perp}\,\hp_z\,\hat\vz
\,,
\ee
where $\hp_{i}$ are direction cosines of the relative momentum $\vp$. 
Surface scattering leads to pair-breaking and suppression of the normal component of the order parameter, 
$\Delta_{\perp}<\Delta_{\parallel}$. 
For weak confinement ($D\gg D_c$) $\Delta_{\parallel}, \Delta_{\perp}\rightarrow \Delta$ away from the boundaries.
But, for strong confinement the B-phase is anisotropic with $\Delta_{\perp}<\Delta_{\parallel}$ everywhere.
Surface scattering results in multiple Andreev reflections and the formation of a spectrum of Fermionic states
confined on the surface. A local description of the surface spectrum and order parameter can be obtained from
solutions of the Bogoliubov equations,
\be
\cHb\left(\nicefrac{\hbar}{i}\,\grad,\vR\right)\ket{\varPsi} = \varepsilon\ket{\varPsi}
\,,
\ee
where $\ket{\varPsi}$ is a four-component Nambu spinor wavefunction for an energy eigenstate of \He\ in the confined
geometry.
\section{Quasiclassical Formulation}\label{sec-Eilenberger}

For \He\ the ratio of the Fermi wavelength, $\hbar/p_f\approx 1\,\mbox{\AA}$, to the size of Cooper pairs,
$\xi_0=\hbar v_f/2\pi k_{\text{B}}T_c\approx 800\,\mbox{\AA}$, is the basis of the quasi-classical approximation to
the Bogoliubov equation.\cite{ser83}
The expansion is achieved by factoring the fast- and slow spatial variations of the spinor wave function,
$\ket{\varPsi} = e^{i\vp\cdot\vR/\hbar}\,\ket{\varPsi_{\vp}}$,
and retaining leading order terms in $\hbar/p_f\xi_0\ll 1$, which yields Andreev's equation,\cite{and64}
\be\label{Andreev_Equation}
\left(\varepsilon\tauz - \whDelta(\vp,\vR)\right)
\,\ket{\varPsi_{\vp}}
+
i\hbar\,\vv_{\vp}\cdot\grad\,\ket{\varPsi_{\vp}}
= 0
\,,
\ee
where $\vp=p_f\hat{p}$ is the Fermi momentum and $\vv_{\vp}=v_f\hat{p}$ is the Fermi velocity. The latter defines classical
trajectories for the propagation of Bogoliubov excitations, which are coherent superpositions of normal-state particles and
holes, with amplitudes given by the Andreev-Nambu spinor, $\ket{\varPsi_{\vp}}$.

Gorkov's propagator is the Greens function for the Bogoliubov equation. In the quasiclassical limit the causal (retarded in
time) propagator, $\whGR(\vp,\vR,\varepsilon)$, is a $4\times4$ matrix, whose diagonal components in particle-hole space,
$\hmfg$, $\hmfg'$, are $2\times2$ spin matrix quasi-particle (hole) propagators, while the off-diagonal components, $\hmff$
and $\hmff'$, are the propagators for Cooper pairs. The full structure of the quasiclassical propagator can be expressed in
terms of spin scalar and vector components of the quasi-particle (quasi-hole) propagators, $\g$ and $\vec{\g}$ ($\g'$ and
$\vec{\g}'$), and spin-singlet and spin-triplet components of the pair propagator (conjugate pair propagator), $\f$ and
$\vec\f$ ($\f'$ and $\vec\f'$),
\be\label{eq-quasiclassical_propagator}
\whGR = 
\begin{pmatrix}
\gR + \gvecR\cdot\vec\vsigma 
& 
\fR\,i\sigma_y + \fvecR\cdot(i\vec\vsigma\sigma _y) 
\\
\fpR\,i\sigma_y + \fvecpR\cdot(i\sigma_y\vec\vsigma) 
& 
\gpR + \gvecpR\cdot\vec{\sigma}^{\text{tr}}
\end{pmatrix}
\ee
The propagator obeys Eilenberger's transport equation
\be\label{eq-Eilenberger_Equation}
\commutator{\epsR\tauz - \whDelta(\vp,\vR)}{\whGR} + i\hbar\vv_{\vp}\cdot\grad\whGR = 0
\,,
\ee
which is supplemented by the normalization condition,\cite{eil68}
\be\label{eq-normalization_condition}
[\whGR(\vp,\vR,\varepsilon)]^2 = -\pi^2\,\widehat{1}
\,,
\ee
with $\epsR=\varepsilon + i0^{+}$ and boundary conditions for the propagators defined on classical trajectories that scatter
off surfaces or interfaces.\cite{kur87}
\subsection{Propagator for {\Heb} near a specular surface}

%-----------------------------------------------------------------------------------------------------------
\begin{figure}[t]
\includegraphics[width=0.4\textwidth]{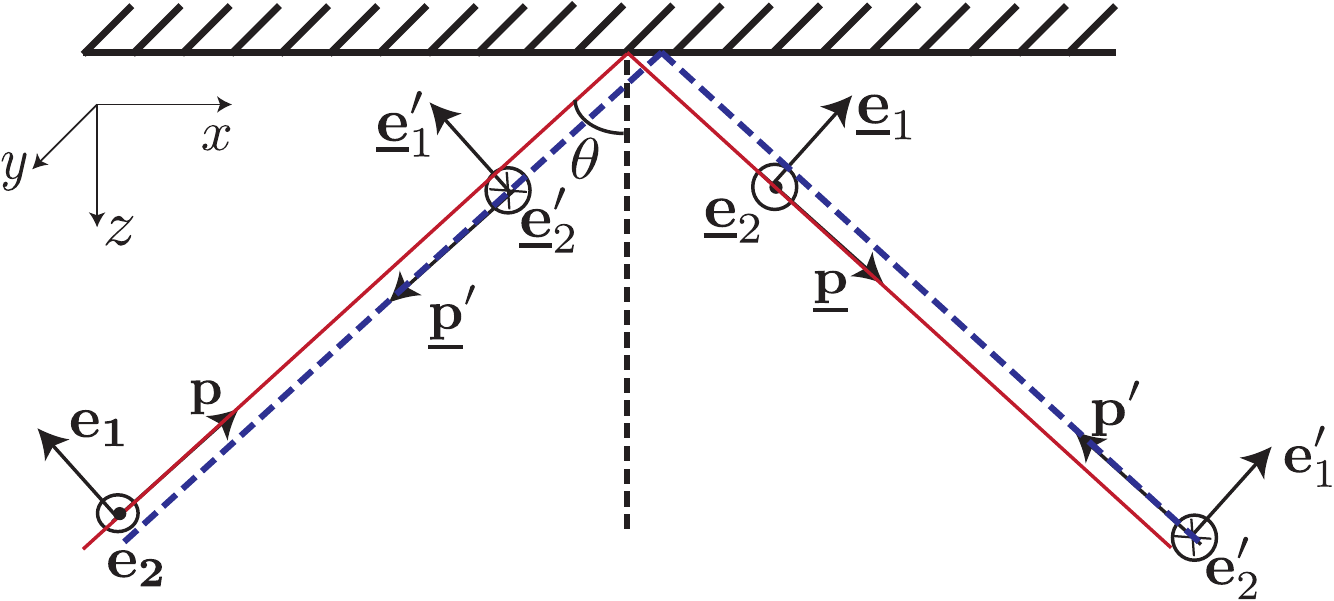}
\caption{The incoming (outgoing) trajectory is represented by $\bf{p}$ ($\underline{\bf{p}}$); $\bf{e}_1$ and $\bf{e}_2$ are
directions transverse to $\bf{p}$, with $\bf{e}_2 \equiv \hat{\vz}\times\hat{\vp}$. The angle between incident
trajectory and surface normal is $\theta$. The coordinates for the time-reversed trajectory pair
($\bf{p}'$,$\underline{\bf{p}}'$) are denoted by primes.}
\label{fig-trajectories}
\centering
\end{figure}
%-----------------------------------------------------------------------------------------------------------

Consider a specularly reflecting surface located at $z=0$ bounding superfluid \Heb\ in the half space, $z > 0$.
In order to investigate the spectrum of fermion excitations, as well as local magnetic and flow properties of
confined \He, we solve Eilenberger's transport equation along classical trajectories in the vicinity of the
boundary at $z=0$ as shown in Fig. \ref{fig-trajectories}.
For a specularly reflecting surface that is far from any other boundary, only a single
reflection: $\vp\rightarrow\underline{\vp}$ couples the propagators for incoming $(\vp)$ and outgoing
$(\underline{\vp})$ trajectories. 

Using the scalar and vector representation for the matrix propagator in Eq. \ref{eq-quasiclassical_propagator} we can
transform Eq. \ref{eq-Eilenberger_Equation} into blocks of coupled equations for the components. This transformation is
described in Appendix \ref{appendix-equation_blocks}.
For \Heb\ in zero magnetic field and zero flow the solution for the propagators components for a deformed B-phase
defined by Eq. \ref{eq-d_vector_B-phase} reduce to scalar and vector quasi-particle propagators and three
spin-triplet Cooper pair propagators; the spin-singlet propagator vanishes:
\ber
\gR 
    &=& -\frac{\pi\epsR}{\lambda(\varepsilon)}
	\left[
		1 - \frac{\Delta_{\perp}^2\cos^2\theta}{(\epsR)^2 - \Delta_{\parallel}^2\sin^2\theta}\,
        e^{-2\lambda(\varepsilon)\,z/\hbar\,v_z}
	\right]
\,,
\label{eq-normal_propagator_scalar}
\\
\gvecR 
       &=& \pi
           \left[
           \frac{\Delta_{\perp}\Delta_{\parallel}\sin\theta\cos\theta}
                   {(\epsR)^2 - \Delta_{\parallel}^2 \sin^2\theta}
           \right]\,
        e^{-2\lambda(\varepsilon)\,z/v_z}\,\,\ve_2
\,,
\hspace*{16mm}
\label{eq-normal_propagator_vector}
\\
\fR_z  &=& -
         \pi
         \frac{\Delta_{\perp}\cos\theta}{\lambda(\varepsilon)}
         \Bigg[1 - e^{-2\lambda(\varepsilon)\,z/\hbar\,v_z}\Bigg]
\nonumber\\
        &&+ 
         \pi
         \Bigg[
         \frac{\epsR\,\Delta_{\perp}\cos\theta}{(\epsR)^2 - \Delta_{\parallel}^2\sin^2\theta}
         \Bigg]\,\,
         e^{-2\lambda(\varepsilon)\,z/\hbar\,v_z}
\,,
\label{eq-anomalous_propagator_z}
\\
\fR_{\parallel} 
         &=& +
          \pi
          \frac{\Delta_{\parallel}\sin\theta}{\lambda(\varepsilon)} 
          \left[1 - \frac{\Delta_{\perp}^2\cos^2\theta}{(\epsR)^2 - \Delta_{\parallel}^2\sin^2\theta}\, 
           e^{-2\lambda(\varepsilon)\,z/\hbar\,v_z} 
          \right]
\,,
\label{eq-anomalous_propagator_x}
\hspace*{8mm}
\eer
where $-\pi/2\le\theta\le\pi/2$ is the angle between the incoming trajectory, $\vp$, and the normal to the surface,
$\hat\vz$, $\ve_1$ and $\ve_2$ are unit vectors transverse to the incoming trajectory, with $\ve_2$ being
perpendicular to the plane containing $\vp$ and $\vp_\parallel$, as shown in Fig. \ref{fig-trajectories}.
These results are obtained for the planar deformed B-phase order parameter defined by the $\vec{d}$-vector in 
Eq. \ref{eq-d_vector_B-phase}. 
The exponential in Eqs. \ref{eq-normal_propagator_scalar}-\ref{eq-anomalous_propagator_x} depends on the projection of the
group velocity along the normal to the interface, $v_z\equiv v_f\cos\theta$, and the function $\lambda(\varepsilon)$ defined
as
\ber\label{eq-tsuneto}
\lambda(\varepsilon) &=& \lim_{\eta\rightarrow 0^{+}}\sqrt{|\Delta(\vp)|^2 - ({\varepsilon+i\eta})^2}
\nonumber\\
                     &=& \sqrt{|\Delta(\vp)|^2 - \varepsilon^2}\,\times\,
                         \Theta(|\Delta(\vp)| - |\varepsilon|)                       
\nonumber\\
                     &+& i\,\sgn(\varepsilon)\sqrt{\varepsilon^2 - |\Delta(\vp)|^2}\,\times\,
                            \Theta(|\varepsilon|-|\Delta(\vp)|)
\,,
\hspace*{10mm}
\eer
where $|\Delta(\vp)|^2\equiv\Delta_{\parallel}^2\sin^2\theta+\Delta_{\perp}^2\cos^2\theta$ defines the continuum
gap edge for the anisotropic B-phase.

The retarded components of the quasiclassical propagator in Eq. \ref{eq-quasiclassical_propagator} have the
following meanings: the scalar component, $\gR$, is related to the local quasiparticle spectral function, while the
spin vector component, $\gvecR$, determines the spin density spectral function. The off-diagonal component in
particle-hole space, $\fvecR$, determines the spectral function for spin-triplet pairing correlations.
The conjugate spin-triplet component is related by symmetry:
$\fvecpR=-\fvecR(-\vp,z,-\varepsilon)^{*}$. For a non-magnetic interface and for zero
magnetic field the spin-singlet pairing correlations vanish, i.e. $\fR=\fpR=0$.
Finally, note that the spin-vector component of the quasiparticle propagator is normal to both $\hat\vz$ and
$\vec\vp_{\parallel}$, while the spin-triplet components are projected onto the film coordinates:
$\vec{\f}^{\text{R}} = \fR_x\,\hat\vx + \fR_y\,\hat\vy + \fR_z\,\hat\vz$, with $\fR_{\parallel}\equiv -(\fR_x +
i\fR_y)\,e^{-i\phi}$.
\subsection{Density of states}\label{sec-DOS}

%-----------------------------------------------------------------------------------------------------------
\begin{figure}[t]
\centering
\includegraphics[width=0.5\textwidth]{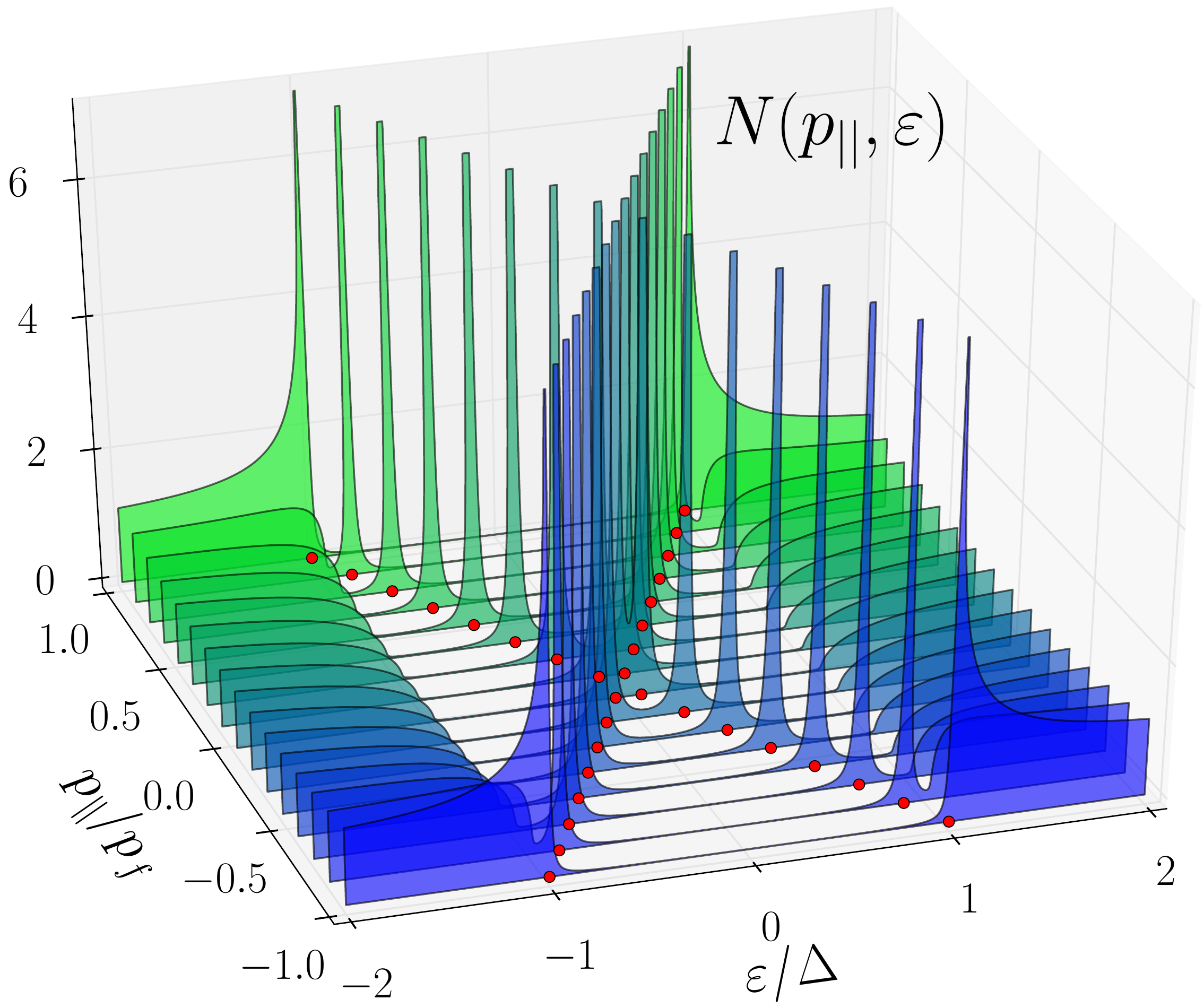}
\caption{Spectral function (LDOS) at a specular boundary ($z=0$) as a function of the 
         in-plane momentum $-p_f \le p_{\parallel} \le p_f$ and energy $\eps/\Delta$. The Fermi level is
         the line $\eps=0$. The red dots denote the energy and momentum of the Andreev bound states.}
\label{fig-LDOS}
\end{figure}
%-----------------------------------------------------------------------------------------------------------

The quasiparticle spectral function, or momentum-resolved local density of states (LDOS), is given by the imaginary
part of the quasiparticle propagator,
\be
N(\vp, z, \varepsilon) = -\frac{1}{\pi} \mathfrak{Im}\,\gR(\vp,z,\varepsilon)
\,,
\ee
where $\gR$ is the scalar contribution to the retarded Greens function in Eq. \ref{eq-normal_propagator_scalar}.
The LDOS is in units of the single-spin, normal-state DOS at the Fermi energy, $N_f$, and
consists of two contributions: the continuum spectrum defined by $|\varepsilon|>|\Delta(\vp)|$ and the
surface Andreev bound-state spectrum for energies below the continuum $|\varepsilon|<|\Delta(\vp)|$.
The surface bound state spectrum is reflected in the poles of $\gR$. There are positive and negative energy
branches of Fermionic states with dispersion relations corresponding to a pair of Dirac cones,
\be\label{eq-Dirac_Dispersion}
\eps_{\pm}^{\text{B}}(\vp_{\parallel}) = \pm\,c\,|\vp_{\parallel}|
\,,
\ee
where $\vp_{\parallel}$ is the in-plane momentum of the Fermions confined on the surface, and
$c=\Delta_{\parallel}/p_f$ is their velocity. The bound-state contribution to the LDOS is given by
\ber\label{eq-DOS_Majorana}
N^{\text{B}}(\vp,z,\eps) 
&=& \frac{\pi\Delta_{\perp}\cos\theta}{2}\,e^{-2\Delta_{\perp}\,z/\hbar\,v_f}
\nonumber\\
&\times&
\left[\delta(\eps-\eps_{+}^{\text{B}}(\vp_{\parallel}))+\delta(\eps - \eps_{-}^{\text{B}}(\vp_{\parallel}))\right]						 
\,.
\eer
The Andreev bound states are confined to the surface within a distance defined by the length scale 
\be
\xi_{\Delta} = \hbar v_f/2\Delta_{\perp}
\,.
\ee
Under strong confinement, $\Delta_{\perp}\rightarrow 0$, the Andreev states penetrate deep into the bulk,
however, their spectral weight also vanishes. 

There are two contributions to the contiuum spectrum. The first term in Eq. \ref{eq-normal_propagator_scalar} gives the bulk
contribution to the LDOS,
$N^{\text{bulk}}(\vp,\varepsilon) = \Theta(|\varepsilon| - |\Delta|)\,|\varepsilon|/\sqrt{\varepsilon^2 - |\Delta|^2}$.
In addition there is a surface-induced contribution to continuum spectrum obtained from $\Im\lambda(\eps)$ in the spatially
varying term of Eq. \ref{eq-normal_propagator_scalar}. The full LDOS for $|\eps|>|\Delta|$ is then given by
\ber
N^{\text{C}}(\vp,\,z,\eps) 
&=& 
\frac{\abs{\eps}}{\sqrt{\eps^2-|\Delta|^2}}
\\
&\times&\hspace*{-3mm}
\left[
	1 - \left(\frac{\Delta_{\perp}^2\cos^2\theta}{\eps^2 - \Delta_{\parallel}^2\sin^2\theta}\right)
	    \cos\left(\frac{2\,z\,\sqrt{\eps^2-|\Delta|^2} }{\hbar\,v_f\cos\theta}\right)
\right].
\nonumber
\eer
Note that the divergence in the LDOS for $|\eps|\rightarrow|\Delta|$ of the bulk spectrum is converted to a square root
threshold for $z=0$, with the transfer in spectral weight appearing in the bound state spectrum. The full LDOS is shown in
Fig. \ref{fig-LDOS}. The bound state energies (red dots) disperse linearly with the in-plane momentum $p_{\parallel}$. The
continuum spectrum is shown for energies above (below) $|\Delta|$ ($-|\Delta|$).
\subsection{Pair-breaking}\label{sec-Gap}

\begin{figure}[t]
\hspace*{-5mm}
\includegraphics[width=0.45\textwidth]{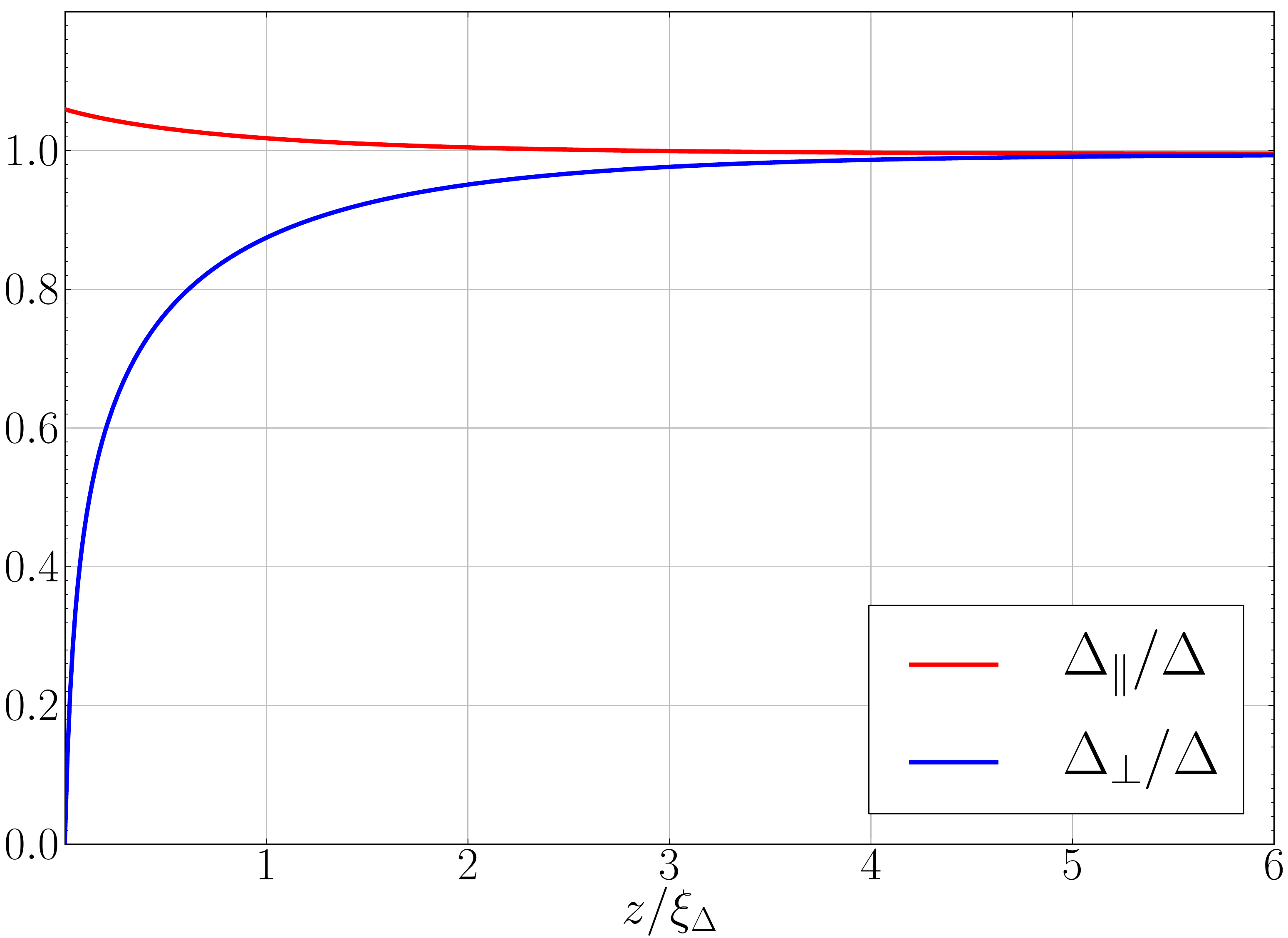}
\caption{Spatial variation of the B-phase order parameter near a specularly reflecting boundary. 
The order parameter components vary on the scale of $\xi_{\Delta}$.}
\label{fig-gap_functions}
\end{figure}

The spectral density for spin-triplet pairing correlations is
\be
\vec{\cP}(\vp,z,\varepsilon) = -\frac{1}{\pi}\Im\,\vec{\f}^{\text{R}}(\vp,z,\varepsilon)
\,,
\ee
also in units of $N_f$, determines the components of the mean pair potentials, $\Delta_{\perp}(z)$ and
$\Delta_{\parallel}(z)$, via the self-consistency condition, i.e. the weak-coupling BCS ``gap equation'',
\ber
\vec{d}(\vp,z) = \langle
						v(\vp,\vp')\,
						\int^{+\Omega_{\text{c}}}_{-\Omega_{\text{c}}}\hspace*{-2mm}d\varepsilon
						\tanh\left(\frac{\varepsilon}{2T}\right)
						\vec\cP(\vp',z,\varepsilon)
						\rangle_{\vp'}
\,,
\eer
where $\Omega_c\ll E_F$ is the bandwidth of attraction for the spin-triplet, p-wave pairing interaction,
$v(\vp,\vp')=3v_1\,\hat\vp\cdot\hat\vp'$, which is integrated over the occupied states defining the pair spectrum
and averaged over the Fermi surface, $\langle\ldots\rangle_{\vp'}\equiv\int\,d\Omega_{\vp'}/4\pi(\ldots)$.
The pairing interaction, $v_1>0$, and cutoff, $\Omega_c$, are eliminated in favor of the $T=0$ bulk gap,
$\Delta=2\Omega\,e^{-1/v_1}$, or the transition temperature, $k_{\text{B}}T_c=1.13\Omega_c\,e^{-1/v_1}$.
Integration over the thermally occupied spectrum can be transformed to a sum over Matsubara energies,
$\varepsilon_n=(2n+1)\pi T$, using the analyticity of $\vec\fR$ in the upper half of the complex energy plane.
Projecting out the normal ($z$) and in-plane ($x,y$) components of $\vec{d}(\vp,z)$ yields the gap equations,
\ber
\Delta_{\perp}     &=& 3v_1\int\dangle{p'}\,p'_z\,T\sum_{\varepsilon_n}\f_z(\vp,z,\varepsilon_n)
\,,
\\
\Delta_{\parallel} &=& 3v_1\int\dangle{p'}\,p'_x\,T\sum_{\varepsilon_n}\f_x(\vp,z,\varepsilon_n)
\,,
\eer
where $\f_z$ ($\f_{\parallel}$) is obtained from Eq. \ref{eq-anomalous_propagator_z}
(\ref{eq-anomalous_propagator_x}) by analytic continuation: 
$\varepsilon\rightarrow\,i\varepsilon_n$ with 
$\lambda(\varepsilon)\rightarrow\sqrt{\varepsilon_n^2 + |\Delta(\vp)|^2}$,
and $\f_x =\f_{\parallel}\cos\phi$ where $\phi$ is the azimuthal angle of incident trajectory. 
The results for $\Delta_{\perp,\parallel}(z)$ are shown in Fig. \ref{fig-gap_functions} for $T=0.2T_c$. 
The normal component of the order parameter, $\Delta_{\perp}(z)$ is suppressed to zero at the boundary. The strong
pair-breaking of the $\hat{p}_z$ component of the mean-field order parameter is explicit in Eq.
\ref{eq-anomalous_propagator_z}: the even frequency pairing correlations vanish at the boundary, while the
odd-frequency pairing correlations do not contribute to the mean field order parameter.
By contrast, the in-plane components are weakly enhanced at the boundary compared to the bulk value. 
The enhancement originates from the bound-state contribution to the in-plane pairing correlations 
that is explicit in Eq. \ref{eq-anomalous_propagator_x}.
Both components converge to the bulk gap on a scale set by the bound-state confinement length, 
$\xi_{\Delta}$. 
Exact gap profiles for the B-phase order can be obtained by computing the propagators with updated values of
the order parameter until self consistency is achieved.\cite{zha87}

%-----------------------------------------------------------------------------------------------------------
%-----------------------------------------------------------------------------------------------------------
\subsection{Projection Operators and Majorana Fermions}\label{sec-Projection}

The normalization condition (Eq. \ref{eq-normalization_condition}) for the quasiclassical Greens function provides
a sum rule on the spectral weight shared between continuum states for Fermions and Cooper Pairs, as well as 
Andreev bound states.
The normalization condition can also be used to extract information on the internal structure (particle-hole
coherence) of the states identified in the LDOS. In particular, Shelankov showed that the Nambu
matrix propagator, $\whmfG^{\text{R}}$, is related to projection operators,\cite{she80}
\be
\widehat{\mathbb P}^{\text{R}}_{\pm} \equiv \frac{1}{2}\left[ \widehat{1}\pm\frac{i}{\pi}
\whmfG^{\text{R}}(\vp,\,z,\eps)\right]
\,,
\ee
for Fermionic quasi-particles ($+$) and quasi-holes ($-$). 
The normalization condition implies that $\widehat{\mathbb P}^{\text{R}}_{\pm}$ obey the identities
for projection operators in Nambu space,
\ber
(\widehat{\mathbb P}^{\text{R}}_{\pm})^2 = \widehat{\mathbb P}^{\text{R}}_{\pm}
\,,\quad
\widehat{\mathbb P}^{\text{R}}_{+}\widehat{\mathbb P}^{\text{R}}_{-} = 
\widehat{\mathbb P}^{\text{R}}_{-}\widehat{\mathbb P}^{\text{R}}_{+} = 0
\,,\quad
\widehat{\mathbb P}^{\text{R}}_{+}+\widehat{\mathbb P}^{\text{R}}_{-} = \widehat{1}
\,.
\eer
In the normal-state, $\whmfG^{\text{R}}_{\text{N}}=-i\pi\tauz$, and we immediately obtain the projection operators in Nambu
space for normal-state quasi-particle, $\widehat{\mathbb P}^{\text{R}}_{+}=\nicefrac{1}{2}(\widehat{1}+\tauz)$, and quasi-hole
states, $\widehat{\mathbb P}^{\text{R}}_{-}=\nicefrac{1}{2}(\widehat{1}-\tauz)$.
The remarkable feature is that Shelankov's operators retain their interpretation as projection operators even for
inhomogeneous superfluid condensates.

Consider the projection of a normal-state quasi-particle with spin $\ket{\vz\uparrow}$ onto 
the $4\times4$ Nambu space defined by the B-phase projectors in the vicinity of a surface,
\be
\ket{\varPsi} = \widehat{\mathbb P}^{\text{R}}_{+}\,
\begin{pmatrix} 1 \\ 0 \\ 0 \\ 0 \end{pmatrix}
              = 
\frac{1}{2}
\begin{pmatrix}
(1+\frac{i}{\pi}\g^{\text{R}}) 								\\
\frac{i}{\pi}(\g_x^{\text{R}} + i\g_y^{\text{R}})			\\
\frac{i}{\pi}(\f_x^{\text{R}} + i\f_y^{\text{R}})			\\
-\frac{i}{\pi}\,\f_z^{\text{R}}
\end{pmatrix}
\,.
\ee
The azimuthal angle, $\phi$, of the in-plane momentum, $\vp_{\parallel}$, factors out of the chiral components: 
\be
\g_x^{\text{R}}+i\g_y^{\text{R}} 
= \g_{\parallel}^{\text{R}}\times(-i e^{i\phi}), \quad \quad \f_x^{\text{R}}+i\f_y^{\text{R}} 
= \f_\parallel^{\text{R}} \times(-e^{i\phi})
\,.
\ee
For the surface bound states, the projected state is evaluated by
integrating over an infinitesimal bandwidth around the delta function at the bound-state energy.
Thus, we obtain two branches of Fermionic bound states corresponding 
$\eps^{\text{B}}_{+}(\vp_{\parallel})$ and $\eps^{\text{B}}_{-}(\vp_{\parallel})$,
\be\label{eq-Majorana}
\ket{\varPsi^{(\pm)}(\vp_{\parallel})} = u(\theta,z)\,
                    \Big[e^{-i\phi/2}\ket{\Phi_{+}} \mp e^{+i\phi/2}\ket{\Phi_{-}}\Big]
\ee
where the amplitude is
\be\label{eq-majorana_amplitude}
u(\theta,z) = \frac{\pi}{4}\Delta_{\perp}\cos\theta\,e^{-z\,/\xi_{\Delta}}
\,,
\ee
and the Nambu spinors, $\ket{\Phi_\pm}$, are given by
\be
\ket{\Phi_+} = \begin{pmatrix} 	1	\\	0	\\	0	\\	-i	\end{pmatrix}
\,,
\quad
\ket{\Phi_-} = \begin{pmatrix}	0	\\	+i	\\	1	\\	0	\end{pmatrix}
\,.
\ee
Note that $\ket{\Phi_{+}}$ is an equal amplitude superposition of a normal-state quasi-particle with spin
$\ket{\vz\uparrow}$, and a normal-state quasi-hole with spin $\ket{\vz\downarrow}$, while 
$\ket{\Phi_{-}}$ is an equal amplitude superposition of a normal-state quasi-particle with spin
$\ket{\vz\downarrow}$, and a normal-state quasi-hole with spin $\ket{\vz\uparrow}$. 
These two spinors are eigenvectors of the Nambu spin operator,
\be
\widehat{S}_z = \frac{\hbar}{2}
\begin{pmatrix} \vsigma_z	& 	0 	\\	0	&	\vsigma_y\vsigma_z\vsigma_y \end{pmatrix}
\,,
\ee
with
\be
\widehat{S}_z \ket{\Phi_{\pm}} = \pm\frac{\hbar}{2}\,\ket{\Phi_{\pm}}
\ee

Thus, for the negative energy bound state, which is fully occupied at $T=0$, then for any 
momentum eigenstate with $0\le p_{\parallel}\le p_f$ and $\phi=0$ the state is 
described by the Nambu spinor,
\be\label{eq-Majorana_zero}
\ket{\varPsi_{-}(\phi=0)} \sim \ket{\Phi_{+}} + \ket{\Phi_{-}}
\,,
\ee
which is the equal amplitude particle-hole spinor with $S_y=+\hbar/2$, i.e. the spin is polarized along the $+\vy$
direction for all $\vp_{\parallel} || +\vx$.
Similarly, for $\phi=\pi/2$ we have 
\be\label{eq-Majorana_pi2}
\ket{\varPsi_{-}(\phi=\pi/2)} \sim \ket{\Phi_{+}} - \ket{\Phi_{-}}
\,,
\ee
which describes equal amplitude particle-hole states with $S_x=-\hbar/2$ for $\vp_{\parallel} || +\vy$.
The spinors $\ket{\Psi^{\pm}}$ obtained from the Nambu propagator describe Majorana Fermions confined on the surface of \Heb\
for any value of $\vp_{\parallel}$, and are equivalent to Majorana spinors obtained by Nagai et al.\cite{nag09} and
Mizushima\cite{miz12} directly from solutions of the Bogoliubov equations. This construction shows that the Majorana property
of the surface Andreev bound states is encoded in the Nambu propagator, and thus the spectral functions for the spin and mass
currents.

\section{Helical Spin Current}\label{sec-Spin_Current}

The correlation between the spin projection and momentum of the
negative energy surface states is encoded in Eqs. \ref{eq-Majorana}-\ref{eq-Majorana_pi2} for the Nambu spinors.  
Occupation of the negative energy states at $T=0$ leads to a ground-state helical spin-current. This is a
key signature of a 3D TRI topological superfluid.
The existence of a ground-state spin current, with spin polarization transverse to $\vp$, can also be inferred from Eq.
\ref{eq-normal_propagator_vector} for the spin-vector component of the quasiparticle propagator, which depends explicitly on
the spectrum of Majorana Fermions.
\subsection{Spin current spectral function}\label{sec-Spin_Current_DOS}

The spin density spectral function is a vector in spin space that provides the contribution to the spin density from states
with energies in the interval ($\varepsilon$, $\varepsilon+d\varepsilon$). This spectral function, in units of
$N_f\,\hbar/2$, is derived from the vector component of the quasiparticle propagator in Eq. \ref{eq-normal_propagator_vector},
\be
\vec{S}(\vp, z, \eps) = -\frac{1}{\pi} \mathfrak{Im}\,\gvecR(\vp, z, \eps)
\,.
\ee 
The vector propagator, $\gvecR(\vp, z; \eps)$, has the same pole as the scalar propagator corresponding to
the spectrum of Majorana Fermions on the surface. The two branches of surface states carry oppositely oriented spin
polarization with equal spectral weight,
\be\label{eq-spectral_spin-bound}
\vec{S}^{\text{B}} = \frac{\pi\Delta_{\perp}\cos\theta}{2}
\negthickspace
           \left[\delta(\eps-\eps^{\text{B}}_{-}(\vp_{\parallel}))
\negmedspace - \negmedspace    
\delta(\eps-\eps^{\text{B}}_{+}(\vp_{\parallel}))
           \right]
\negthinspace
            e^{-2\Delta_{\bot} z/\hbar\,v_f}\,{\bf e}_{2}
\,.
\ee
Furthermore, the spin polarization of surface excitations are opposite for any pair of time-reversed trajectories, $\vp$ and
$\vp'= - \vp$. This implies a net spin current confined on the surface at any temperature $T<T_c$.
We define the spin-current spectral density as the local spin-current density of states for a pair of time-reversed trajectories,
\be\label{eq-spectral_spin-current}
\vJ_\alpha (\vp, z, \eps) = 2N_f\,\times\,\frac{\hbar}{2} \vv_{\vp} 
\left[S_\alpha(\vp, z; \eps) - S_\alpha (\vp', z; \eps)\right]
\,,
\ee
where $\vJ_\alpha$ is the $\alpha$ spin component of spin current flowing in the $\vv_{\vp}$ direction, $N_f$ is
the normal-state density of states at Fermi level for one spin, and $\vp'$ denotes the time-reversed trajectory of
$\vp$, with $\vv_{\vp'} = - \vv_{\vp}$.
The spin-current density is then obtained by thermally occupying this spectrum and integrating over all incident
trajectories,
\be\label{eq-spin-current}
\vec{\vJ}(z) = \int_{\text{in}}\dangle{\vp}\int^{+\infty}_{-\infty}d\eps\,f(\eps)\,\vec{\vJ}(\vp, z, \eps)
\,,
\ee
where $f(\eps) = 1/(e^{\beta\eps} + 1)$ is the Fermi function and $\beta=1/T$. Note that $\vec{\vJ}(z)$ is a tensor under 
$\spinorbit$ rotations, and can be represented by a $3\times 3$ matrix with components, $\mbox{J}_{\alpha i}(z)$.
%-----------------------------------------------------------------------------------------------------------
\begin{figure}[t]
\centering
\includegraphics[width =0.45\textwidth]{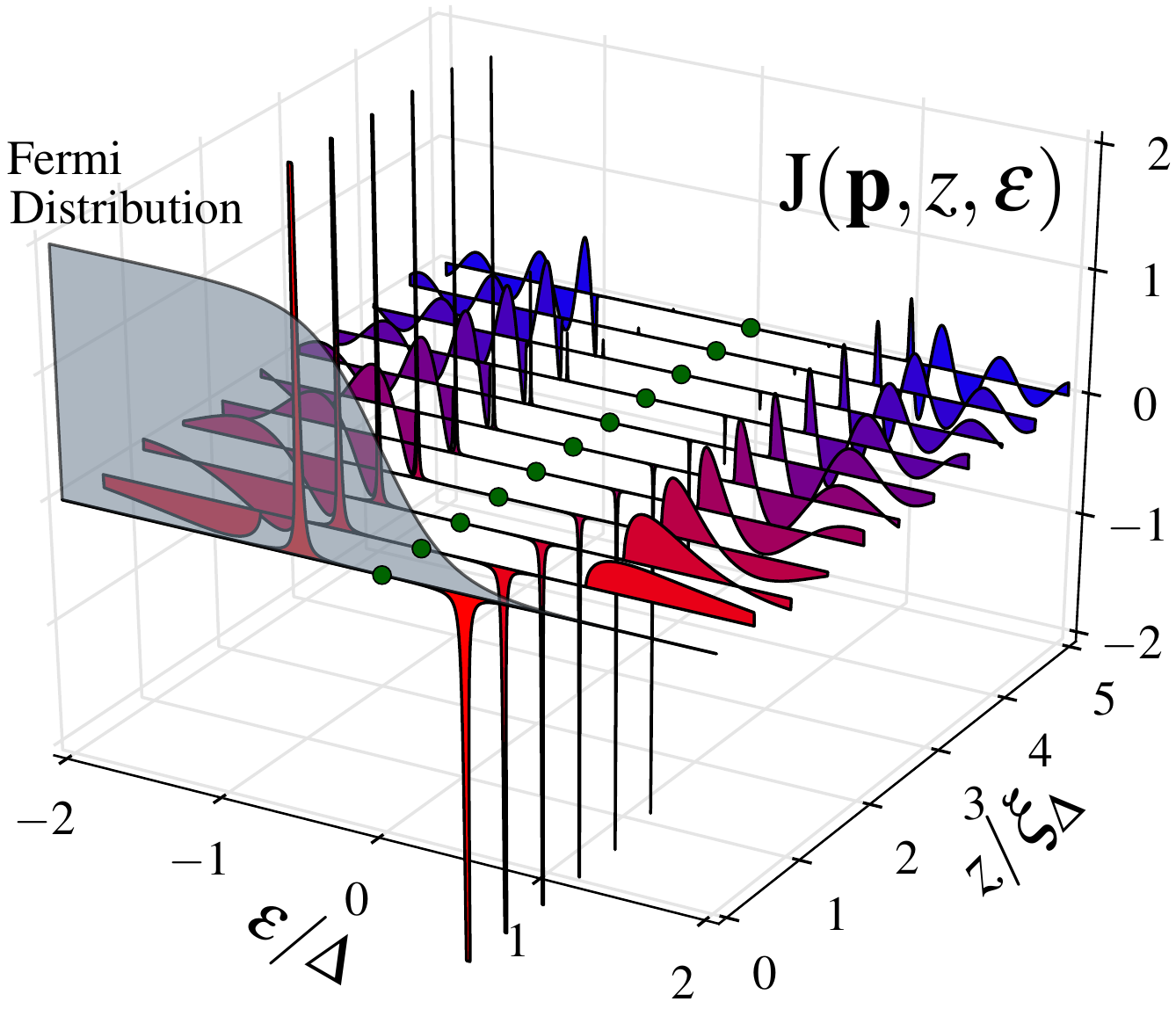}
\caption{Spectral function for the surface spin current, $\vec{\vJ}(\vp,z,\eps)$, vs. depth, $z$, for trajectory angle, $\theta=\pi/6$ (in-plane momentum $p_{\parallel}=p_f/2$). The gray region represents the thermally
occupied states at low, but finite, temperature. Tomasch oscillations develop for $z>0$.}
\label{fig-JDOS}
\end{figure}
%-----------------------------------------------------------------------------------------------------------
\subsection{Ground-state surface spin current}\label{sec-Spin_Current_Ground-State}

\begin{figure}[t]
\centering
\includegraphics[width=0.35\textwidth]{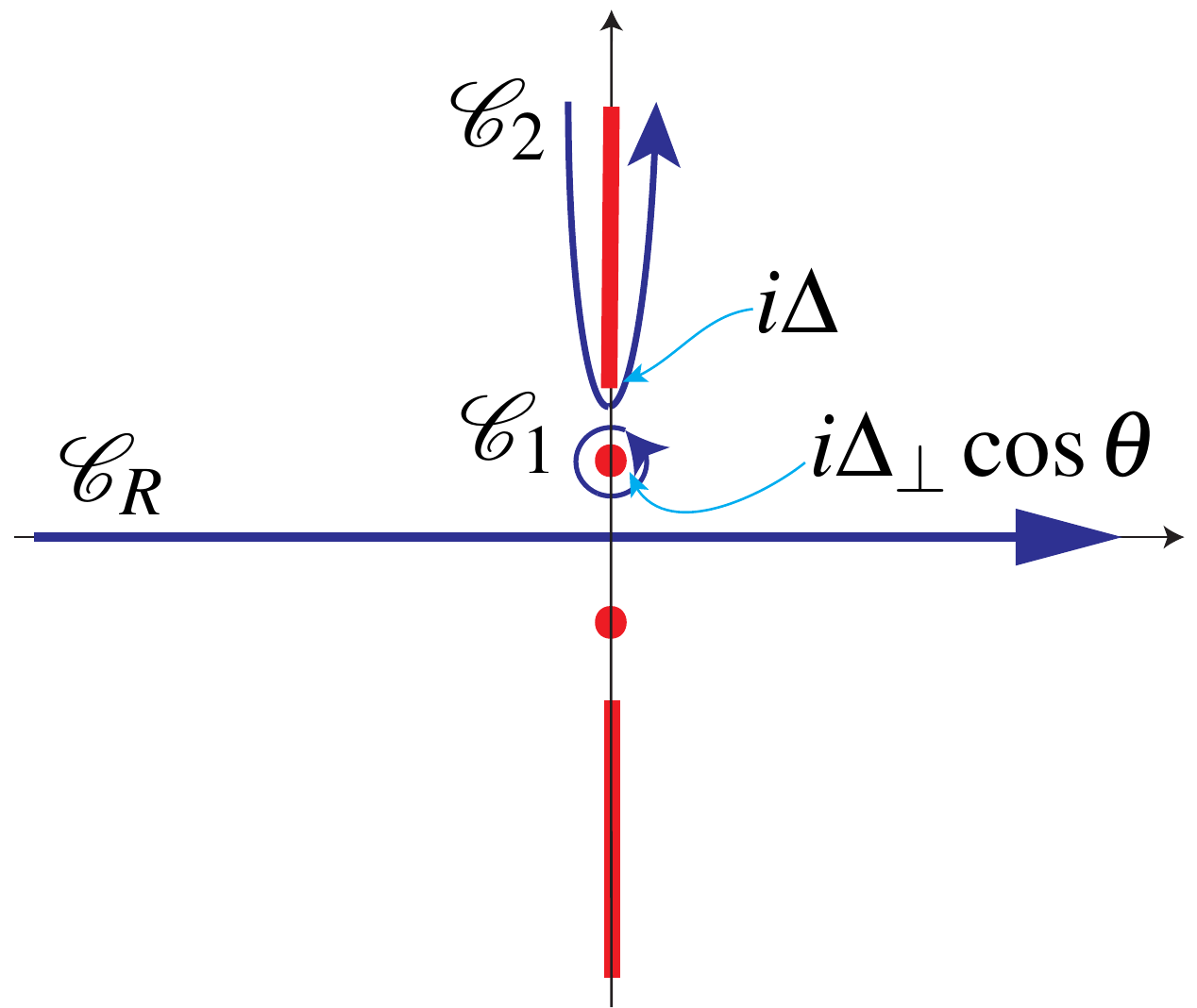}
\caption{Singularities defining the continuum contribution to the ground-state spin-current in Eq.
\ref{eq-spin-current_continuum-kernel2}. Integration along the real axis, $\cC_{\text{R}}$, is transformed to integration
around the resonance pole, $\cC_{1}$, at $\xi=+i\Delta_{\perp}\cos\theta$, plus the branch cut, $\cC_{2}$, extending from
$+i\Delta$ along the imaginary axis.}
\label{fig-countour_integration}
\end{figure}

For the ground state only negative energy states are occupied, however there are contributions to the spin current from
both the surface bound states and the continuum. The bound state contribution to the spin-current density is obtained by
evaluating Eq. \ref{eq-spin-current} with the bound-state spectral density obtained from Eqs. \ref{eq-spectral_spin-bound}
and \ref{eq-spectral_spin-current}. The matrix representation is then,
\be\label{eq-spin-current_bound-state}
\vec{\vJ}^{\text{B}}(z)
=
\frac{\hbar}{2}\frac{\pi N_f v_f \Delta_{\perp}}{6}
\begin{pmatrix}
0	&	-1	&	0\\
+1	&	0	&	0\\
0	&	0	&	0
\end{pmatrix}
\,e^{-2\Delta_{\perp} z/\hbar\,v_f}
\,.
\ee
Note that for fixed $z$ the spin-current is a spatially homogeneous, in-plane current of spins, also aligned in-plane and
normal to the direction of flow, i.e. the current flowing along $+\hat\vx$ carries spins polarized along $+\hat\vy$, while
the current flowing along $+\hat\vy$ carries spins polarized along $-\hat\vx$. This is a \emph{helical} spin-current in
which the current flowing along a direction $\hat{\ve}$ in the $xy$ plane transports spin polarized along
$\hat\vz\times\hat\ve$.
It is also clear from Eqs. \ref{eq-normal_propagator_vector} and \ref{eq-tsuneto} that the negative energy continuum
spectrum contributes to the ground-state spin current. This is a ``response'' of the condensate to formation of
the surface Majorana spectrum. 
Bound-state and continuum contributions to the spin-current are shown clearly in Fig. \ref{fig-JDOS} for the magnitude of
the spin-current spectral function at $p_{\parallel}=p_f/2$.
In particular, the continuum contribution to the spin density spectral function becomes,
\be\label{eq-spectral_spin-continuum}
\vec{S}^{\text{C}}=\frac{\Delta_{\bot} \Delta_{\parallel}  \sin\theta \cos\theta}
              {\eps^2-\Delta_{\parallel}^2\sin^2\theta}\sgn(\eps) \sin(2\sqrt{\eps^2-\Delta^2}\,z/ \hbar\,v_z)\,\ve_{2}
\,,
\ee
and the corresponding continuum contribution to the spin-current density can be expressed as
\be\label{eq-spin-current_continuum}
\mbox{J}^{\text{C}}_{\alpha i}(z)
=
-\frac{N_f \hbar v_f}{2\pi}\int_{\text{in}} d\Omega_{\vp}\,\Delta_{\bot}\Delta_{\parallel}\sin\theta\cos\theta\,
\ve_{2}^{\alpha}\,\hat{p}_i\,
\mbox{I}(\vp)
\,,
\ee
where the $\mbox{I}(\vp)$ is defined by as an integration over the occupied negative energy continuum,
\be\label{eq-spin-current_continuum-kernel}
\mbox{I}(\vp)
=
\int_{-\infty}^{-\Delta}\,
\frac{d\eps}{\eps^2-\Delta_{\parallel}^2\sin^2\theta}\sin\left(2\sqrt{\eps^2-\Delta^2}\,z/\hbar\,v_z\right)
\,.
\ee
Note that even though the spin-density spectral function exhibits Tomasch oscillations into the bulk of the condensate, the
continuum contribution to the spin-current is confined to the surface on the scale of $\xi_{\Delta}$.
The integration over the spectrum in Eq. \ref{eq-spin-current_continuum-kernel} is evaluated by extending the energy
integration to positive and negative energies, and transforming to an integration over the radial momentum, or equivalently,
$\xi = v_f p = \sqrt{\eps^2-\Delta^2}$. Thus, we can write Eq. \ref{eq-spin-current_continuum-kernel} as 
\be\label{eq-spin-current_continuum-kernel2}
\hspace*{-3mm}
\mbox{I}(\vp)=\frac{1}{2}\Im\,\oint_{\cC_{\text{R}}} d\xi\, 
\frac{\xi}{\sqrt{\xi^2+\Delta^2}}\frac{1}{\xi^2+\Delta_{\bot}^2\cos^2\theta}\,e^{2i\xi z/\hbar\,v_z}
\,,
\ee
where $\cC_{\text{R}}$ is the real axis. The integrand has simple poles on the imaginary axis at 
$\pm i\Delta_{\perp}\cos\theta$ and branch cuts at $[\pm i\Delta,\pm i\infty]$ as shown in Fig. 
\ref{fig-countour_integration}.
The integral along the real axis is
transformed to integrals around the pole at $\cC_1$ and around the branch cut from $+i\, \Delta$ to $+i\,\infty$:
$\mbox{I}(\vp)_{\mathcal{C}_R}=\mbox{I}_{\mathcal{C}_1}(\vp)+\mbox{I}_{\mathcal{C}_2}(\vp)$.
%
%-----------------------------------------------------------------------------------------------------------
%-----------------------------------------------------------------------------------------------------------
%
The resonance pole at $\xi=+i\Delta_{\perp}\cos\theta$ gives a contribution to the ground-state spin-current that
\emph{exactly cancels} the bound-state contribution to the spin-current from Eq. \ref{eq-spin-current_bound-state}. 
Thus, the net spin-current is determined by the the non-resonant contribution $\mbox{I}_{\cC_{2}}(\vp)$ from the branch cut,
which evaluates to 
\be\label{eq-spin-current_kernel_non-resonant}
\mbox{I}_{\mathcal{C}_2}(\vp)=-\int_0^\infty\frac{d\eps}{\eps^2+\Delta_{\parallel}^2 
\sin^2\theta}e^{-2\sqrt{\Delta^2+\eps^2}\,z/\hbar\,v_z}
\,,
\ee
which shows explicitly the response of the continuum states originating from the (off-resonant) surface bound-state.
The cancellation of the spin-current carried by the surface bound-state by a resonant contribution from the negative energy
continuum is essentially identical to the cancellation of mass current carried by the chiral edge states from a similar
resonance term for the continuum states of 2D \Hea.\cite{sau11}
The resulting ground-state spin current density is then given by Eq. \ref{eq-spin-current_continuum} evaluated with Eq.
\ref{eq-spin-current_kernel_non-resonant}. While it is clear from Eq. \ref{eq-spin-current_kernel_non-resonant} that the
spin-current is confined to the surface, there is no single confinement scale as was the case for the bound-state
contribution in Eq. \ref{eq-spin-current_bound-state}.
However, we can evaluate the \emph{sheet} spin-current by integrating the spin-current density from the surface into the
bulk. The resulting sheet spin-current confined on the surface of \Heb\ is given by
\be\label{eq-sheet_spin-current_ground-state}
\vec{\vK} = \int_{0}^{\infty}dz\,\,\vec{\vJ}(z) 
 =
\mbox{K}(0)\,
\bpm
0	&	-1	&	0	\\
+1	&	0	&	0	\\
0	&	0	&	0
\epm
\,,
\ee
with the magnitude given by
\ber
\mbox{K}(0)
&=&
N_f v_f^2\hbar^2\,\Delta_{\perp}\Delta_{\parallel}
                  \int_{\text{in}}\frac{d\Omega_\vp}{4\pi}\, \sin\theta \cos^2\theta \sin^2\phi 
\nonumber\\
&\times&
\int_0^\infty d\eps \, \frac{1}{(\varepsilon^2 + \Delta^2_\parallel\sin^2\theta)\sqrt{\Delta^2+\varepsilon^2}}
\label{eq-sheet_spin-current_energy_integral}
\\
& = & 
I_{\mbox{\footnotesize K}}\left(\frac{\Delta_{\parallel}}{\Delta_{\perp}}\right)\,\times\,N_f v_f^2\hbar^2
\,,
\eer
where the function $I_{\mbox{\footnotesize K}}(x)$ is given in Appendix \ref{Appendix-Sheet_Spin-Current}. 
For weak confinement, $\Delta_\perp = \Delta_\parallel$, we obtain a ground-state spin current
\be\label{eq-ground_state_spin-current_weak}
\mbox{K}(0)\equiv\nicefrac{1}{18}\,N_f v_f^2\hbar^2 = \nicefrac{1}{6}\,n_{\text{2D}}\,v_f\,\nicefrac{\hbar}{2}
\,,
\ee
that is independent of the pairing energy scale $\Delta$, and depends only on the spin-current carried by a normal-state
quasiparticle, $v_f\,\nicefrac{\hbar}{2}$, and the areal density of \He\ atoms, $n_{\text{2D}}\equiv n\hbar/p_f$.
This result agrees with that of Tsutusmi and Machida\cite{tsu12b} if we neglect the Fermi-liquid correction to effective
mass in the expression $N_fv_f^2=\nicefrac{3}{2}n/m^{*}$ in Eq. \ref{eq-ground_state_spin-current_weak}, i.e. if we set
$m^{*}= m_{\text{3}}$, the atomic mass of \He.

However, for strong confinement, $\Delta_{\perp}\ll\Delta_{\parallel}$, the universality is destroyed,
$\mbox{K}(0)\approx\,\nicefrac{1}{4}
\left(\Delta_{\perp}/\Delta_{\parallel}\right)^3\,n_{\text{2D}}\,v_f\,\nicefrac{\hbar}{2}$, which vanishes for
$\Delta_{\perp}=0$.
This is expected as there is no Andreev bound-state or spin-current on the surface of the planar phase of \He, as is clear
from Eq. \ref{eq-majorana_amplitude} for the Majorana amplitude.
\subsection{Temperature dependence of the spin current}\label{sec-Spin_Current_vs_T}

%-----------------------------------------------------------------------------------------------------------
\begin{figure}[t]
\includegraphics[width=0.45\textwidth]{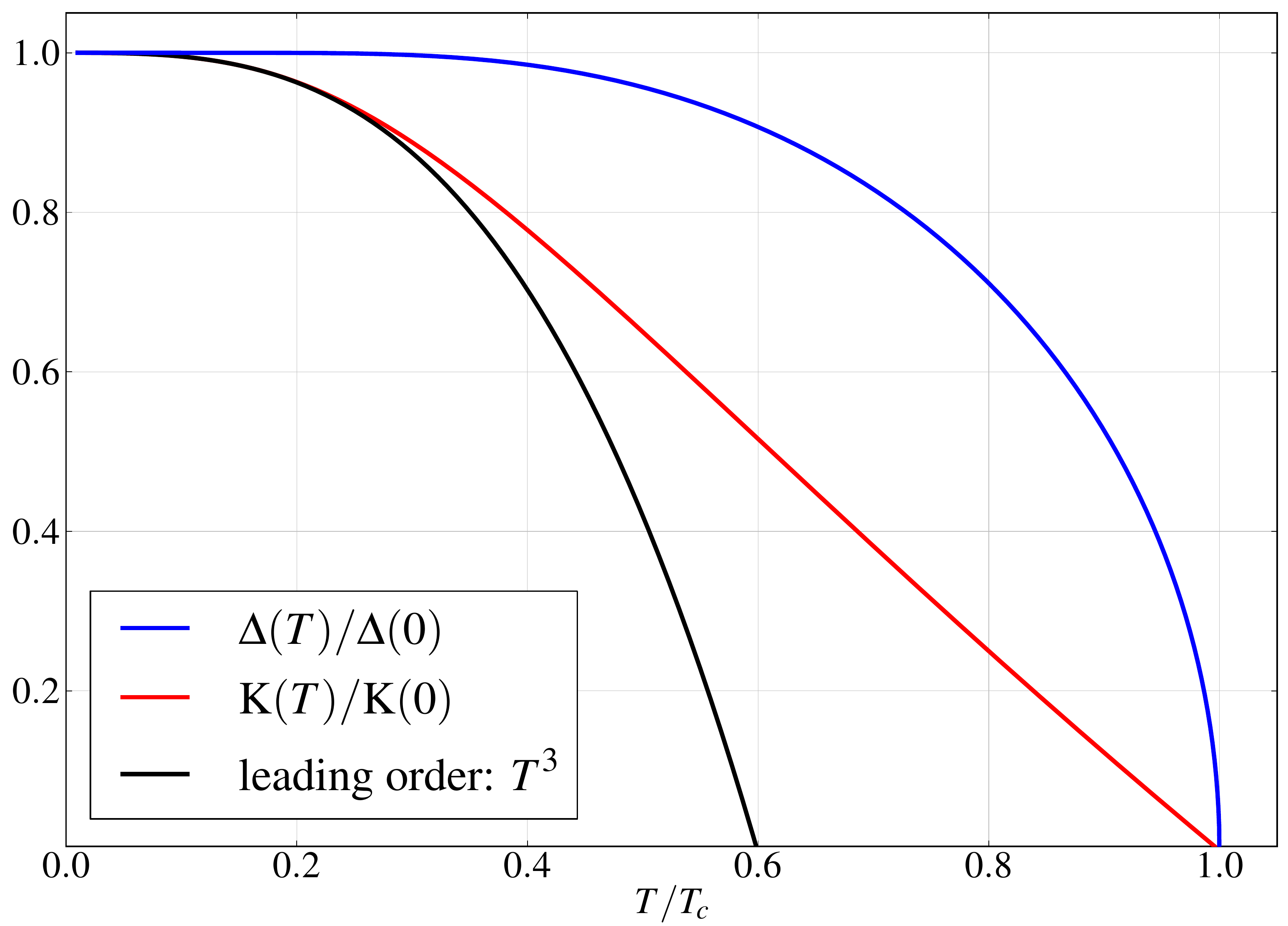}
\caption{(color online) The temperature dependence of sheet spin current, $\mbox{K}(T)$, calculated for the isotropic
B-phase ($\Delta_{\perp}=\Delta_{\parallel}$), is shown in red. The leading order correction, $\propto -T^3$, to the
ground-state spin current is shown in black. The temperature dependence of the bulk B-phase gap $\Delta(T)$ is shown in
blue.}
\label{fig-sheet_spin-current_vs_T}
\end{figure}
%-----------------------------------------------------------------------------------------------------------

The analysis of the ground-state spectral current also applies at finite temperature, leading to the following result for 
the kernel of Eq. \ref{eq-spin-current_continuum} for the temperature dependent spin current density,
\be\label{eq-spin-current_continuum-kernel_T}
\hspace*{-3mm}
\mbox{I}(\vp)=\frac{1}{2}\Im\oint_{\mathcal{C}_2}\negmedspace d\xi\negmedspace
\frac{\xi\tanh\left(\sqrt{\xi^2+\Delta^2}/2T\right)}{\left(\xi^2+\Delta_{\perp}^2\cos^2\theta\right)
\sqrt{\xi^2+\Delta^2}}\,e^{2i\xi z/\hbar\,v_z}
.
\ee
The thermal distribution transforms the branch cut into a sum over pole contributions at $\xi_n = i\sqrt{\eps_n^2 +
\Delta^2}$, where $\eps_n=(2n+1)\pi T$ for $n=0,\pm 1,\pm 2,\ldots$ are the Fermion Matsubara energies.\footnote{We use the
Matsubara expansion: $\frac{1}{2\eps}\tanh\left(\frac{\eps}{2T}\right)=T\sum_{\eps_m} \frac{1}{\eps^2+\eps_m^2}$.}
The resulting kernel is given by
\be
\hspace*{-3mm}
\mbox{I}(\vp)=-\pi T\sum_{\eps_n}\frac{1}{\eps_{n}^2+\Delta_{\parallel}^2\sin^2\theta}\,
               e^{-2\sqrt{\Delta^2+\eps_{n}^2}\,z/\hbar\,v_z}
\,.
\ee
We can then express the sheet current at finite temperature as
\ber\label{eq-sheet_spin-current_vs_T}
\vec{\vK}(T)
&=& 2 
N_f v_f^2\hbar^2\,\Delta_{\perp}\Delta_{\parallel}\int_{\text{in}}\frac{d\Omega_{\vp}}{4\pi}\,
\sin\theta\cos^2\theta\,(\ve_{2}\,\hat{p})
\nonumber\\
&\times& 
\pi T\sum_{\eps_{n}}\frac{1}{\left(\eps_{n}^2+\Delta_{\parallel}^2\sin^2\theta\right)\sqrt{\eps_{n}^2+\Delta^2}}
\\
&\equiv&
\mbox{K}(T)\,
\bpm
0	&	-1	&	0	\\
+1	&	0	&	0	\\
0	&	0	&	0
\epm
\,,
\eer
which is easily evaluated numerically for any temperature.
Figure \ref{fig-sheet_spin-current_vs_T} shows that the sheet current, $\mbox{K}(T)$, decreases rapidly at finite
temperatures compared to the bulk B-phase gap. This is a key signature of the Majorana spectrum. At low but finite
temperatures, $T\ll\Delta$, thermal excitation of the positive energy Majorana branch, which carries oppositely directed spin
current to the negative energy bound states (see Fig. \ref{fig-JDOS}), leads to a reduction of sheet spin current from the
ground state. In particular, the leading order correction to the ground state spin current is $\mbox{K}(T)-\mbox{K}(0)\propto
- T^3$, which dominates thermal excitations above the continuum gap of the bulk B-phase.

The leading order low temperature correction is obtained by transforming the Matsubara sum in Eq.
\ref{eq-sheet_spin-current_vs_T} to an integral over real energies,
\ber\label{eq-SK-sheet_current_vs_T}
S_{\mbox{\footnotesize K}}(T)
&=& 
T\sum_{\eps_{n}}\frac{1}{(\eps_{n}^2+\Delta_{\parallel}^2\sin^2\theta)\sqrt{\Delta^2+\eps_{n}^2}}
\\
&=&
\frac{1}{2\pi i}\oint_{\mathcal{C}_u +\mathcal{C}_l}\negthickspace dz\,f(z)
\frac{1}{(z^2-\Delta_{\parallel}^2\sin^2\theta)\sqrt{\Delta^2-z^2}}
\,,
\quad
\eer
where the contours $\cC_{u,l}$ enclose the poles, $\{i\eps_{n}|n=0,\pm 1,\ldots\}$, of the Fermi function 
$f(z)=(1+e^{\beta z})^{-1}$.
The rest of the integrand has poles on the real axis at the Majorana branches, $z=\pm \Delta_{\parallel}\sin\theta$, and 
branch cuts for the continuum spectrum at $[\mp\infty\,,\,\mp\Delta]$. 
Transforming the integration to the real axis we obtain contributions from the Majorana branches,
\be\label{eq-SK-sheet_current_vs_T-Majorana}
S_{\mbox{\footnotesize K}}^{\text{Majorana}} = 
\frac{\tanh\left(\Delta_{\parallel}\sin\theta/2T\right)}
     {2\Delta_{\perp}\Delta_{\parallel}\sin\theta\cos\theta}
\,,
\ee
and the fully gapped continuum states, which can be expressed as a single integral for both positive energy excitations
and de-population of the negative energy continuum,
\be\label{eq-SK-sheet_current_vs_T-gapped}
S_{\mbox{\footnotesize K}}^{\text{gapped}} 
= 
\frac{1}{\pi}\int_{\Delta}^{\infty}\,d\eps
\frac{\tanh\left(\eps/2T\right)}{(\eps^2-\Delta_{\parallel}^2\sin^2\theta)\sqrt{\eps^2-\Delta^2}}
\,.
\ee

First note that the $T=0$ result for the sheet current is obtained from the leading term in the Euler-Maclaurin expansion 
of Eq. \ref{eq-SK-sheet_current_vs_T}, i.e. $\pi S_{\mbox{\footnotesize K}}(0)$ reduces to the 
integration over the spectrum in Eq. \ref{eq-sheet_spin-current_energy_integral}. 
Alternatively, we take the limit $T\rightarrow 0$ in Eqs. \ref{eq-SK-sheet_current_vs_T-Majorana} and
\ref{eq-SK-sheet_current_vs_T-gapped} to obtain 
$S_{\mbox{\footnotesize K}}(0)=S_{\mbox{\footnotesize K}}^{\text{Majorana}}(0)+S_{\mbox{\footnotesize K}}^{\text{gapped}}(0)$.
The leading order correction for finite temperatures is obtained from an expansion of 
$S_{\mbox{\footnotesize K}}(T)-S_{\mbox{\footnotesize K}}(0)$.
For the continuum branches we transform Eq. \ref{eq-SK-sheet_current_vs_T-gapped} to obtain
\ber
&&
S_{\mbox{\footnotesize K}}^{\text{gapped}}(T)
\negthickspace-\negthickspace
S_{\mbox{\footnotesize K}}^{\text{gapped}}(0) =
\hspace*{5.2cm}
\nonumber\\
&&-\frac{2}{\pi}\int_{0}^{\infty}d\xi\,
\frac{1}{\xi^2+\Delta^2_\perp\cos^2\theta}\frac{1}{\sqrt{\xi^2+\Delta^2}}\frac{1}{1 + e^{\sqrt{\xi^2 + \Delta^2}/T}}
\,.
\eer
The spectrum is gapped provided $\Delta_{\perp}\ne 0$, and the correction from the continuum spectrum is exponentially small,
$\mbox{K}^{\text{gapped}}(T)\negthickspace-\negthickspace\mbox{K}^{\text{gapped}}(0) \sim
-\mbox{K}^{\text{gapped}}(0)\,e^{-\Delta_{\perp}/T}$ for $T\ll\Delta_{\perp}$.
However, for the Majorana branches
\be
\hspace*{-3mm}
S_{\mbox{\footnotesize K}}^{\text{Majorana}}(T)
\negthickspace-\negthickspace
S_{\mbox{\footnotesize K}}^{\text{Majorana}}(0)
=
\frac{-1}{\Delta_\perp \Delta_\parallel|\sin\theta|\cos\theta}\frac{1}{1 + e^ {\Delta_\parallel|\sin\theta|/T}}
\,.
\ee
The resulting temperature dependence of the sheet spin-current arising from the Majorana spectrum is given by 
\ber
\hspace*{-3mm}
\mbox{K}^{\text{Majorana}}(T)\negthickspace-\negthickspace\mbox{K}^{\text{Majorana}}(0)
&\negthickspace=\negthickspace&
-\frac{\pi N_f v^2_f\hbar^2}{4} 
\left(\frac{T}{\Delta_\parallel}\right)^3
\negthickspace\negthickspace\int_0^{\Delta_\parallel/T}\negthickspace\negthickspace dx \,\frac{x^2}{1 + e^x} 
\nonumber
\\
&\negthickspace\approx\negthickspace&
-\frac{3\pi}{8}\zeta(3)\,N_f v^2_f\hbar^2\left( \frac{T}{\Delta_\parallel} \right )^3 
\,,
\label{eq-SK-sheet_current_vs_T-Majorana-leading_order}
\eer
where the second line gives the leading order correction for $T\ll\Delta_\parallel$.
Thus, the leading order correction to the sheet current is obtained from thermal excitation of the Majorana branches.
For weak confinement ($\Delta_{\perp}=\Delta_{\parallel}=\Delta$) we obtain
\be
\mbox{K}(T) \approx \mbox{K}(0)\left(1 -\frac{27\pi}{4}\zeta(3)\,\left(\frac{T}{\Delta}\right)^3 \right)
\,,
\ee
which is the black curve plotted in Fig. \ref{fig-sheet_spin-current_vs_T} and compared with the numerical result for all $T$
from Eq. \ref{eq-sheet_spin-current_vs_T}. Also note that the coefficient of the $T^3$ correction in Eq.
\ref{eq-SK-sheet_current_vs_T-Majorana-leading_order} is consistent with the bounds obtained in Ref. \onlinecite{tsu12b}.
This temperature dependence of the surface spin-current is a direct signature of the Majorana excitations. An
experimental probe of the equilibrium spin current would provide a key signature of the Majorana surface spectrum. 
Alternatively, there are well known experimental probes to measure the temperature dependence of the mass flow through a
``superleak''.

\section{Mass current in a  channel}\label{sec-Mass_Current}

\begin{figure}[t]
\includegraphics[width=0.4\textwidth]{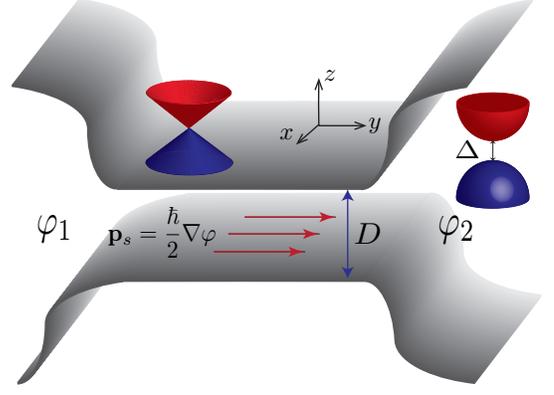}
\caption{The bulk B-phase is gapped, while the surface spectrum is a gapless Dirac cone. Mass flow is 
         represented by the pair momentum, $\vp_s=\nicefrac{1}{2}\,\hbar\grad\varphi$, in a channel 
         of width $D > D_c \approx 9\xi_0$.}
\label{fig-channel}
\centering
\end{figure}

Consider the response of the surface and bulk spectrum to a phase bias between two bulk reservoirs of \Heb\
as shown in Fig. \ref{fig-channel}.
The bias will generate a phase gradient along the channel which we express in terms of the Cooper pair momentum,
$\vp_s\equiv\nicefrac{\hbar}{2}\grad\varphi$, or ``flow field'', and which we take to be directed along the $\hat{y}$ axis
of the channel.
Andreev bound states are present on both surfaces of the channel. In zero flow these are gapless Majorana branches.
However, the spectrum is expected to be modified by the flow field. 

%-----------------------------------------------------------------------------------------------------------
%-----------------------------------------------------------------------------------------------------------

\subsection{Gauge transformation and Doppler shift}\label{sec-Mass_Current_Symmetries}

The flow field, $\vp_s=\frac{\hbar}{2}\grad\varphi(\vR)$, is generated by spatial variations of the phase of the
order parameter,
\be
\whDelta_{\varphi}(\vp,\vR)=
\begin{pmatrix}		0											& 	\hDelta(\vp,\vR)\,e^{i\varphi(\vR)}	\cr
				\hDelta(\vp,\vR)^{\dagger}\,e^{-i\varphi(\vR)}	&		0
\end{pmatrix}
\,,
\ee
where $\hDelta(\vp,\vR)$ is the order parameter in the absence of flow.
A local gauge transformation defined by
\be
\widehat{\cU} = \exp{\left(\frac{i}{2}\varphi(\vR)\tauz\right)}
\,,
\ee
``removes'' the phase of the order parameter
\be\label{eq-local_gauge_transformation}
\whDelta =  \widehat{\cU}^{\dagger}\,\,\whDelta_{\varphi}\widehat{\cU}
\,,
\ee
and transforms the Bogoliubov-Nambu Hamiltonian to
\be
\cHb_{\varphi}\xrightarrow[]{\cU} \cHb' = \cHb + \vp_s\cdot\vv(\vp)\widehat{1}
\,,
\ee
where $\cHb$ is the Hamiltonian in the absence of the flow field (Eq. \ref{eq-Hamiltonian_B-phase}) and
$\vp_{s}\cdot\vv(\vp)$ is the Doppler shift from the condensate flow. 
The Hamiltonian, including the Doppler shift, again anti-commutes with the conjugation symmetry, 
\be
\charge\,\cHb'\,\charge^{\dag} = -\cHb - \vp_s\cdot\vv(\vp)\,\widehat{1} = -\cHb'
\,,
\ee
but the flow field breaks time-reversal symmetry, defined as complex conjugation ($\conjugate$) combined with a $\pi$ rotation
in spin space $\time=i\sigma_y\conjugate$,
\be
\time\,\cHb'\,\time^{-1} = \cHb - \vp_s\cdot\vv(\vp)\,\widehat{1} \ne \cHb'
\,.
\ee
Given the role that $\time$-symmetry plays in defining the topological winding number for \Heb\ in Eq.
\ref{eq-winding_number-3D},\cite{qi09,vol09,miz12} the question is: if $\time$-symmetry is broken will
a gap develop in the surface spectrum?
In the case of broken $\time$-symmetry by a magnetic field a gap typically develops.\cite{vol10} 
However, for in-plane fields the situation is more subtle. A discrete rotation symmetry combines with time-reversal to protect
the topological invariant, and thus the Majorana spectrum and zero energy state, at least a sufficiently low magnetic
fields.\cite{miz12a}

Similarly, the broken $\time$-symmetry by the flow field can be repaired by a \underline{joint} $\pi$ rotation about the normal to the surface, 
\be
\piz\time\,\cHb'\,\time^{\dag}\piz^{\dag} = \cHb + \vp_s\cdot\vv(\vp)\,\widehat{1} = \cHb'
\,.
\ee
Thus, under the repaired time-reversal and conjugation symmetry, $\Gamma\equiv\piz\times\time\times\charge$,
we have $\Gamma\,\cHb'\,\Gamma^{\dag} = -\cHb'$
The combined symmetry preserves the topological winding number, and thus the Majorana zero mode.\cite{miz12,miz12a}
This result is born out by the calculation of the spectral response to the flow field, which provides an important
signature of the Majorana spectrum.

The unitary transformation to the Doppler shifted Hamiltonian carries over to the quasi-classical transport equation for the
Nambu matrix propagator in the presence of the flow field. In particular, the local gauge transformation used in Eq.
\ref{eq-local_gauge_transformation} applied to the transport equation gives
\be\label{eq-Eilenberger_Equation_Doppler}
\commutator{(\epsR - \vp_s\cdot\vv_{\vp})\tauz - \whDelta(\vp,\vR)}{\whGR} + i\hbar\vv_{\vp}\cdot\grad\whGR = 0
\,,
\ee
where $\whGR(\vp,\vR, \eps)=\widehat{\cU}^{\dagger}\,\,\whGR_{\varphi}\widehat{\cU}$ is the gauge transformed propagator, and
the Fermi velocity, $\vv_{\vp}=v_f\hat{p}$, determines the Doppler shift. The normalization condition in Eq.
\ref{eq-normalization_condition} is unchanged.
Thus, the solutions for the quasiparticle and Cooper pair propagators in the presence of a flow field near a surface are given
by Eqs. \ref{eq-normal_propagator_scalar}-\ref{eq-anomalous_propagator_x} evaluated with the Doppler shifted excitation
energy, $\epsR\rightarrow\epsR-\vp_s\cdot\vv_{\vp}$. 

In particular, the surface bound-state energies are shifted by $\vp_s\cdot\vv_{\vp}=v_f\,p_s\,\hat{p}\cdot\hat{y}$, which
leads to positive and negative Dirac cones,
\ber
\eps_{\pm}^{\text{B}}(\vp) &=& \pm\,c_{\pm}(\phi) p_f|\sin\theta|
\,,
\eer
with anisotropic velocities, $c_{\pm}(\phi)=\Delta_{\parallel}/p_f \pm p_s/m^{*}\,\sin\phi$,
and a zero energy state for zero in-plane momentum. 
Thus, none of the bound-state energies cross the Fermi level for $p_s \lesssim \Delta_{\parallel}/v_f$.
This leads to two important results for mass transport in a channel: (i) the ground-state mass current is unaffected by the
surface bound state and (ii) the anisotropy in the Majorana spectrum is reflected in a power-law temperature
dependence of the mass current at finite temperatures.

\subsection{Mass current spectral density}\label{sec-Mass_Current_DOS}

The mass current spectral density is defined in terms of the net momentum transported by time-reversed pairs of states $(\vp,-\vp)$ in the energy interval $(\eps,\eps+d\eps)$,\cite{rai96}
\be
\vj_{\text{M}}(\vp, z, \eps) = 2 N_f \vp_{\text{M}} \left[N_{+}( \vp, z, \eps) - N_{-}(\vp, z, \eps) \right]
\,
\ee
where $\vp_{\text{M}}\equiv m_{3}\vv_{\vp}$, $N_{+}(\vp,z,\eps)$ is the DOS for quasiparticles co-moving with the flow
field, $\eps_{\text{D}}=\vv_{\vp}\cdot\vp_s > 0$, and $N_{-}(\vp,z,\eps)=N_{+}(-\vp,z,\eps)$ is the DOS for the
counter-moving states, $\eps_{\text{D}}<0$.
The mass current is then obtained by thermally occupying the spectrum of current carrying states and integrating over all the
trajectories,
\be
\vj_{\text{M}}(z) = \int_{\eps_{\text{D}}>0}\dangle\vp\int_{-\infty}^{+\infty} d\eps\,f(\eps)\vj_{\text{M}}(\vp, z, \eps)
\,.
\ee
Note that the full Fermi surface is covered by summing over the co-moving trajectories. This result for the mass current is
valid for Galilean invariant Fermi liquids.\cite{ser83} However, the Doppler shift of the quasiparticle spectrum leads to a
change in the quasiparticle energy due to molecular field interactions with the modified spectrum and thermal distribution of
quasiparticles. The Fermi liquid effect is substantial for \He\ in the case of mass flow, and is discussed and calculated for
confined \Heb\ in Sec. \ref{sec-Fermi-Liquid_Effects} In this and the following section we omit the Fermi liquid interaction,
in which case the Fermi momentum and Fermi velocity are related by $m_3\,v_f = p_f$.

At $T=0$ the negative energy bound states give a net zero contribution to the ground-state mass current. For every pair of
co-moving and counter-moving bound states, either both have positive energies (both un-occupied) or both have negative
energies (both occupied), \underline{and} even though these pairs have different energies, they contribute equal spectral
weight to the ground-state current (see Eq. \ref{eq-DOS_Majorana}).
Thus, the ground-state mass current is given by the occupied continuum states. The difference in co-moving and counter-moving 
contributions can be transformed the following integration over the continuum spectrum,
\ber
\vj_{\text{M}}(z) 
&=& 
2N_f \oint_{\eps_{\text{D}}>0}\dangle\vp\,\vp_{\text{M}}  
\lim_{E_c\rightarrow\infty}\int_{-E_c-\eps_{\text{D}}}^{-E_c+\eps_{\text{D}}}\negthickspace\negthickspace
d\eps\,\frac{-\eps}{\sqrt{\eps^2-\Delta^2}}
\nonumber\\
&\times&
\left(1+\frac{\Delta_{\bot}^2 \cos^2\theta}{\Delta_{\parallel}^2 
\sin^2\theta-\eps^2}\cos\frac{2z\sqrt{\eps^2-\Delta^2}}{v_f\cos\theta} \right)
\,,
\eer
where the limiting procedure enforces the restriction that only the low energy part of the spectrum contributes to the
current. In particular, the Tomasch oscillations average to zero and the bulk DOS term yields the well known result for the
ground-state superfluid density for bulk \Heb,
\be
\vj_{\text{M}}
=
2N_f\oint_{\eps_{\text{D}}>0}\dangle{\vp}\vp_{\text{M}}\left[2\,\vp_s\cdot \vv_\vp\right] 
= 
\frac{2N_f p_f v_f}{3}\vp_s \equiv n\,\vp_s
\,,
\ee
where $n$ is the density of \He\ atoms. 
\subsection{Mass current at finite temperatures}\label{sec-Mass_Current_vs_T}

At finite temperature the mass current is reduced by thermal excitations. The leading order 
contribution to the mass current from the continuum states at finite temperature is essentially that of bulk \He,
\be
\vj^{\text{C}}_{\text{M}}
=
2N_f \oint_{\eps_{\text{D}}>0}\dangle\vp\,\vp_{\text{M}}\left[2\,\vp_s\cdot \vv_\vp\right] 
\left(1 - Y(\vp,T)\right)
\ee
where the temperature dependent term is given by the Yoshida function for the anisotropic B-phase,
\be
\hspace{-5mm}
Y(\vp,T)
\equiv
\frac{1}{2T} \int_{|\Delta(\vp)|}^\infty d\eps \, 
\frac{\eps}{\sqrt{\eps^2-|\Delta(\vp)|^2}}\,\sech^2\left(\frac{\eps}{2T}\right)
\,.
\ee
Angular integration gives,
\be\label{eq-backflow_continuum_bulk}
\vj^{\text{C}}_{\text{M}} = \left(1 - Y_{\parallel}(T)\right)\,n\,\vp_s
\,,
\ee
where 
\be\label{eq-Y_bulk_in-plane}
Y_{\parallel}(T) = \frac{3}{2}\int_{0}^{1}\,d(\cos\theta)\,\sin^2\theta\,\,Y(\vp,T)
\,,
\ee
is the ``normal'' component for flow in the plane of the film,
which in the limit of weak confinement,
$\Delta_{\perp}\rightarrow\Delta_{\parallel}=\Delta$, is Yoshida function for bulk B-phase.

There is also a surface contribution from the continuum states given by
\ber
\hspace*{-10mm}
Y_{\text{S}}^{\text{C}}(\vp,T) 
&=& 
-
\frac{1}{2}\int_{-\infty}^{+\infty}d\xi\,
\frac{\Delta_{\perp}^2\cos^2\theta}{\xi^2+\Delta_{\perp}^2\cos^2\theta}\,
e^{2i\,\xi\,z}/\hbar v_z\,
\nonumber\\
&\,&\times
\frac{1}{2T}\sech^2\left(\frac{\sqrt{\xi^2 + |\Delta(\vp)|^2}}{2T}\right)\,
\,,
\eer
where we use the change of variables $\xi=\sqrt{\eps^2 - |\Delta(\vp)|^2}$ in the surface contribution to the continuum
spectrum in Eq. \ref{eq-normal_propagator_scalar}.
This term is confined near the surface since the oscillations average to zero for $z\gg\xi_{\Delta}$. If we first
average over the film for $D\gg\xi_{\Delta}$,
\be
\nicefrac{1}{D}\int_{0}^{D}dz\,\exp{2i\xi\,z/\hbar v_z} 
\approx
\left(\frac{\pi\hbar v_f\cos\theta}{2D}\right)\,\delta(\xi)
\,,
\ee
we obtain the net contribution from the surface term
\be\label{eq-Y_continuum-surface}
Y_{\text{S}}^{\text{C}}(\vp,T) 
=
-
\frac{\pi\hbar v_f\,\cos\theta}{4D}\frac{1}{2T}\,\sech^2\left(\frac{|\Delta(\vp)|}{2T}\right)
\,.
\ee
This gives a surface correction to the bulk term in Eq. \ref{eq-backflow_continuum_bulk}, which in the weak confinement limit
reduces to 
\be
Y_{\text{S},\parallel}^{\text{C}}(T) 
= 
-
\frac{3\pi}{32}
\frac{\hbar v_f}{D}\frac{1}{2T}\,\sech^2\left(\frac{\Delta}{2T}\right)
\,.
\ee
At low temperatures both continuum corrections to the ground-state mass current are exponentially small, $\propto 
e^{-\Delta/T}$, and the correction from the surface bound-state dominates.

The Doppler shift of the surface bound-state spectrum leads to a power law reduction in the superfluid density that 
reflects the energy difference between co-moving and counter-moving Majorana excitations.
In particular, thermal occupation of the bound-state contribution to the spectral density gives,
\ber\label{eq-jM_bound-state}
\vj^{\text{B}}_{\text{M}}(z)\negmedspace
&=& \negmedspace
2N_f \oint_{\eps_{\text{D}}>0}\dangle\vp\,
\vp_{\text{M}}\,\left(\frac{\pi}{2}\,\Delta_{\perp}\cos\theta\right)\,e^{-2 z \Delta_{\perp}/\hbar v_f}
\nonumber\\
&\times&\negthickspace\negthickspace
\left[
\tanh\left(\frac{\Delta_{\parallel}\sin\theta -\eps_{\text{D}}}{2T}\right)
-
\tanh\left(\frac{\Delta_{\parallel}\sin\theta + \eps_{\text{D}}}{2T}\right)
\right]
.
\hspace*{5mm}
\eer
which reduces to 
\be\label{eq-Y_bound-state}
\vj^{\text{B}}_{\text{M}}(z) 
= -\frac{3\pi}{8}\,\frac{\Delta_{\perp}}{T}\,I(\Delta_{\parallel}/T)\,e^{-2\Delta_{\perp}z/\hbar v_f}\,n\,\vp_s
\,,
\ee
in the low velocity limit, $p_s v_f \ll \Delta_{\parallel}$, where 
\be
I(\Delta_{\parallel}/T) 
= \left(\frac{2T}{\Delta_{\parallel}}\right)^4\int_{0}^{\Delta_{\parallel}/T}dx\,x^3\sech^2(x) 
\,.
\ee
Thus, the leading low temperature ($T\ll\Delta_{\parallel}$) correction to is 
\be
\vj^{\text{B}}_{\text{M}}(z) 
= 
-\frac{27\pi\zeta(3)\Delta_{\perp}}{4\,\Delta_{\parallel}}\,e^{-2\Delta_{\perp} z/\hbar v_f} 
\left(\frac{T}{\Delta_{\parallel}}\right)^3\times n\,\vp_s
\,.
\ee
The average mass current for confined \Heb, including a factor of two for both surfaces, is then
\be
\overline{\vj}_{\text{M}}
=
\left(1 - (Y_{\parallel}(T) + Y_{\text{S},\parallel}(T))\right)\,n\,\vp_s
\,,
\ee
where $Y_{\parallel}(T)$ is the bulk normal fluid fraction (Eq. \ref{eq-Y_bulk_in-plane}) and $Y_{\text{S},\parallel}(T)$ is
the total surface contribution from Eqs. \ref{eq-Y_continuum-surface} and \ref{eq-Y_bound-state},
\ber
\hspace*{-3mm}
Y_{\text{S},\parallel}(T)
\negthickspace = \negthickspace
\frac{3\pi}{4}\frac{\xi_{\Delta}}{D}\,\frac{\Delta_{\perp}}{T}
\negthickspace\int_0^1\negthickspace\negthickspace dx\,x^3\negthickspace
\left[\sech^2\negthickspace\left(\frac{\Delta_\parallel x}{2T}\right) 
\negthickspace-\negthickspace 
\sech^2\negthickspace\left(\frac{\Delta}{2T}\right) \right]
\negthickspace\,.
\eer
Thus, the leading order correction to the the mass current at low temperature becomes
\be\label{eq-T3_mass_current}
\overline{\vj}_{\text{M}} 
= 
\left(
1 
- 
\frac{27\pi\zeta(3)}{2}\frac{\xi_{\Delta}}{D}\,\frac{\Delta_{\perp}}{\Delta_{\parallel}}
\left(\frac{T}{\Delta_{\parallel}} \right)^3
\right) 
\times n\, \vp_s
\,.
\ee
As in the case of the spin current, the reduction $\propto T^{3}$ reflects the linear dispersion and spectral weight of the
Majorana excitations. 

\subsection{Fermi-liquid correction to the mass current}\label{sec-Fermi-Liquid_Effects}

The motion of a \He\ quasiparticle is strongly influenced by its interaction with the medium of low-energy
excitations near the Fermi surface. These Fermi liquid interactions are equally significant below $T_c$ where they give rise
to a large renormalization of the superfluid fraction in bulk \Heb.\cite{leg65,sau81e} Since the surface Majorana excitations
are linear superpositions of normal-state particle and hole excitations, the current carried by the Majorana excitations
will also be renormalized by Fermi liquid interactions. We include this effect by treating the Doppler effect associated with
the condensate flow as an external vector potential that couples to the group velocity of normal-state particles and holes,
\be
\whSig_{\text{flow}}(\vp) = \vp_s\cdot\vv_{\vp}\,\tauz
\,.
\ee
The flow perturbs the equilibrium distribution of quasiparticles and the condensate. This leads to a Fermi-liquid correction
described by Landau's molecular field self-energy,\footnote{For a description of the quasiclassical formulation of
Fermi liquid effects on the properties of superfluid \He\ see Ref. \onlinecite{ser83}.}
\be\label{eq-molecular-field_self-energy}
\whSig_{\text{FL}}(\vp,\vR) = \int\dangle{p'}\,T\sum_{\eps_n}\,A^s(\vp,\vp')\,\g(\vp',\vR,\eps_{n})\,\tauz
\,,
\ee
where $A^s(\vp,\vp')$ is the forward scattering amplitude, which can be expanded in terms of Legendre polynomials,
$A^s(\vp,\vp')=\sum_{l>0} A_l^s\,P_l(\hp\cdot\hp')$. The scattering amplitudes are related to the Fermi liquid parameters
by $A^{s}_{l} = F^{s}_{l}/1 + F^{s}_{l}/2l+1$. Since we are considering interaction effects on the mass current we have
included only the spin-independent interactions.
With the additional self energies, Eilenberger's transport equation, in the Matsubara representation, becomes, 
\be\label{eq-Eilenberger_Equation_Matsubara}
\commutator{i\eps_n\tauz-\whSig_{\text{flow}}-\whSig_{\text{FL}}-\whDelta}{\whmfG} + i\hbar\vv_{\vp}\cdot\grad\whmfG = 0
\,.
\ee

The molecular field self-energy vanishes in the absence of flow since $\g(\vp,\vR,\eps_{n})$ is odd in $\eps_{n}$ (Eq.
\ref{eq-normal_propagator_scalar} with $\epsR\rightarrow i\eps_{n}$). Thus, we can obtain the linear response of confined
\Heb\ to the imposed flow field by expanding the propagator in $v_fp_s \ll \Delta$, $\g=\g^{(0)} + \g^{(1)} + \ldots$. The
zeroth order propagators are given in Eqs. \ref{eq-normal_propagator_scalar}-\ref{eq-anomalous_propagator_x}, and the linear
correction is obtained by perturbation expansion of Eqs. \ref{eq-Eilenberger_Equation_Matsubara} and
\ref{eq-normalization_condition}.

To obtain the mass current in the film we average the molecular field over the width of the film,
\be
\Sigma_{\text{FL}}(\vp) = \frac{1}{D}\int_0^D dz \,\, \Sigma{\text{FL}}(\vp,z)
\,,
\ee
and similarly for the response function, $\g^{(1)}(\vp,z,\eps_{n})$, to obtain
\ber
\hspace*{-3mm}
\bar{\g}^{(1)}(\vp,\eps_{n})
\negthickspace
&\negthickspace=&
\negthickspace\negthickspace
\pi\Sigma_{\mathsf{tot}}
\Bigg\{
\frac{|\Delta(\vp)|^2}{\left(\eps_{n}^2 + |\Delta(\vp)|^2\right)^{3/2}} 
+ 
\frac{\hbar v_f\Delta_{\perp}^2\cos^3\theta}{D} 
\\
\negthickspace\negthickspace&\times&\negthickspace\negthickspace
\frac{3\eps_{n}^4 + \eps_{n}^2|\Delta(\vp)|^2 + \Delta_\parallel^2\sin^2\theta (\eps_{n}^2 - |\Delta(\vp)|^2) }
     {(\eps_{n}^2+|\Delta(\vp)|^2)^2(\eps_{n}^2+\Delta_\parallel^2\sin^2\theta)^2} 
\Bigg\}
\,,
\nonumber
\eer
where $\Sigma_{\mathsf{tot}} = \Sigma_{\text{flow}} + \Sigma_{\text{FL}}$. To linear order in $v_f p_s$ the molecular field
self energy is obtained by evaluating Eq. \ref{eq-molecular-field_self-energy} with the linear response propagator
$\bar{\g}^{(1)}(\vp,\eps_{n})$. Only the odd-parity interactions contribute to the renormalization of the mass current, and
in bulk \Heb\ only the $l=1$ Fermi liquid interaction contributes due to the isotropic bulk gap.\cite{leg75} For confined
\Heb\ higher order odd-parity terms will also contribute to the renormalization of the mass flow, but the dominant
contribution is expected to be from the $l=1$ interaction since $F^{s}_{1}\approx 6 - 15$ is large at all pressures. With
only the $l=1$ interaction channel, we can express $\Sigma_{\text{FL}} = A^{s}_{1}\,\hp\cdot\vX$, where
\be
\vX = \int\dangle{p}\,\hp\,\Phi(\vp)\,
        \left(\vv_{\vp}\cdot\vp_s + A^{s}_{1}\,\hp\cdot\vX\right)
\,,
\ee
and $\Phi(\vp)\equiv T\sum_{\eps_{n}}\delta\bar{\g}^{(1)}/\delta\Sigma_{\mathsf{tot}}$ is the Matsubara representation of the
bulk and surface contributions to the superfluid fraction.
The solution for the vector self energy $\vX$ can then be expressed in terms of bulk and
surface contributions to the normal fluid fraction obtained in Sec. \ref{sec-Mass_Current_vs_T},
\be\label{eq-vector_potential}
\vX = (1 + \nicefrac{1}{3}F^{s}_{1})\,\nicefrac{1}{3}\,v_f\,\vp_s\,\,
\frac{\left(1-(Y_{\parallel}(T) + Y_{\text{S},\parallel}(T))\right)}
     {1 + \nicefrac{1}{3}F^{s}_{1}\,(Y_{\parallel}(T) + Y_{\text{S},\parallel}(T))}
\,.
\ee 
This vector potential also determines the mass current,
\be
\overline{\vj}_{\text{M}} 
= 
2N_f\int\dangle{p}\,\vp_{\text{M}}\,T\sum_{\eps_{n}}\,\g^{(1)}(\vp,\eps_{n}) = 2N_f m_{3} v_f \vX
\,,
\ee
which reduces to
\be\label{eq-mass_current_Fermi-Liquid}
\overline{\vj}_{\text{M}} 
= 
\frac{\left(1-(Y_{\parallel}(T) + Y_{\text{S},\parallel}(T))\right)}
     {1 + \nicefrac{1}{3}F^{s}_{1}\,(Y_{\parallel}(T) + Y_{\text{S},\parallel}(T))}
\,\,n\,\vp_s
\,,
\ee
which is the same functional form as that of bulk \Heb,\cite{leg75} but with the surface continuum and bound-state
corrections to the bulk Yoshida function included. Note that the first factor of $1 + \nicefrac{1}{3}F^{s}_{1}$ in Eq.
\ref{eq-vector_potential} for the vector potential $\vX$ is the effective mass ratio for \He\ obtained from Galilean
invariance of the interactions. This factor renormalizes $m_{3}v_f\rightarrow m^{*}v_f=p_f$, and gaurantees that we recover
ground state current, $n\,\vp_s$, at $T=0$.
In low temperature limit, the $T^3$ power law for the reduction of the mass current from the Majorana excitations is 
preserved, 
\be\label{eq-T3_mass_current_Fermi-Liquid}
\overline{\vj}_{\text{M}} = 
\left(1 - \frac{27\pi\zeta(3)}{2}\frac{\xi_{\Delta}}{D}\,\frac{\Delta_{\perp}}{\Delta_{\parallel}}\,\frac{m^{*}}{m_{3}}
          \left(\frac{T}{\Delta_{\parallel}} \right)^3
\right) 
\times n\, \vp_s
\,,
\ee
but the prefactor is increased by the factor $m^{*}/m_{3}=1+F^{s}_{1}/3$ compared to that in Eq. \ref{eq-T3_mass_current}.

\begin{figure}[t]
\hspace*{-5mm}
\includegraphics[width=0.5\textwidth]{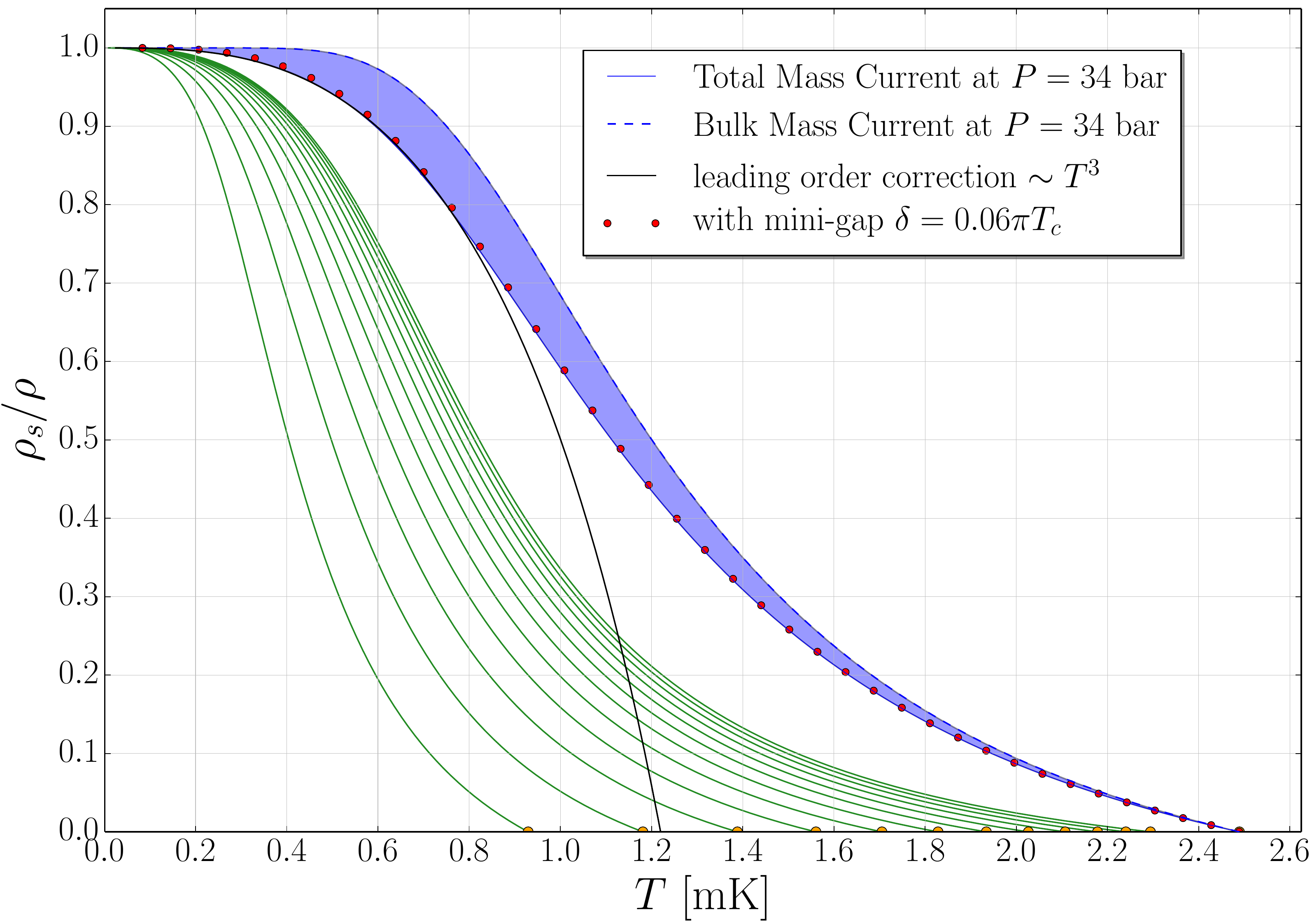}
\caption{(color online) Superfluid mass fraction for a \Heb\ film of width $D = 7.5\, \xi_{\Delta}$ ($D \approx
13.2\,\xi_{0}$) for pressures $p = 0\ldots 22\,\mbox{bar}$ in steps of $2\,\mbox{bar}$ (green) and $p=34\,\mbox{bar}$ (solid
blue). $T_c$ at each pressure is indicated by the orange circles. The bulk superfluid fraction at $34\,\mbox{bar}$ is shown
for comparison (dashed blue), and the shaded region represents the reduction in supercurrent from thermal excitation of the
surface Majorana states. For $T\lesssim 0.9\,\mbox{mK}$ these excitations give a $T^3$ power law (black) for the reduction in
mass current. The effect of the mini-gap, $\delta = 0.06\pi T_c$, in the surface spectrum on the superfluid fraction is shown
for $p=34\,\mbox{bar}$ (red circles).
}
\label{fig-mass_current}
\end{figure}

The temperature dependence of the superfluid fraction for confined \Heb\ calculated from Eq.
\ref{eq-mass_current_Fermi-Liquid} is shown in Fig. \ref{fig-mass_current} for $D=7.5\xi_{\Delta}$ over the full pressure
range of $0-34\,\mbox{bar}$. The pressure dependence of the effective mass and bulk transition temperature were obtained from
the Helium calculator.\cite{har00} For $p=34\,\mbox{bar}$ the superfluid fraction for bulk \Heb\ is included, which
highlights the role of thermally excited surface Majorana Fermions in suppressing the superfluid fraction over the full
temperature range below $T_c$.
For $T\lesssim 0.9\,\mbox{mK}$ these excitations give the $T^3$ power law for the reduction in superfluid mass current
calculated from Eq. \ref{eq-T3_mass_current_Fermi-Liquid}.
Experimental observation of the power law correction from the Dirac spectrum in the fully gapped B-phase would provide direct 
evidence of surface Majorana Fermions.

%-----------------------------------------------------------------------------------------------------------
%-----------------------------------------------------------------------------------------------------------

\subsection{Majorana Fermions on Opposing Surfaces of a Channel}

For well separated surfaces, $D\gg\xi_{\Delta}$, we need only consider single reflections as shown in Fig.
\ref{fig-trajectories} in order to calculate the surface propagator, surface spectral function of gapless Majorana
modes. In particular the Majorana mode with zero energy at $\vp_{\parallel}=0$ corresponds to the reflection
$\vp\rightarrow -\vp$ at normal incidence.
For \Heb\ in a channel of width $D\gtrsim D_c\approx 9\xi_0$ the wave functions for the Majorana modes on opposite surfaces,
which are confined near the surfaces of the channel on the length scale $\xi_{\Delta}=\hbar v_f/\pi\Delta_{\perp}$, are
expected to overlap. This situation typically leads to level splitting and a gap in the otherwise gapless Dirac cone.
Numerical calculations reported in Ref. \onlinecite{tsu11} for \Heb\ confined in a rectangular channel, i.e. confinement in
two directions, suggest that overlap between states confined on opposing walls generates a finite excitation energy, i.e. a
``mini-gap'', in the spectrum at $\vp_{\parallel}=0$, of magnitude $\delta = 0.06\pi T_c$ for a spacing of $D = 13.2
\xi_0$ and $T \ll T_c$.

\begin{figure}[t]
\centering
\includegraphics[width =0.45\textwidth]{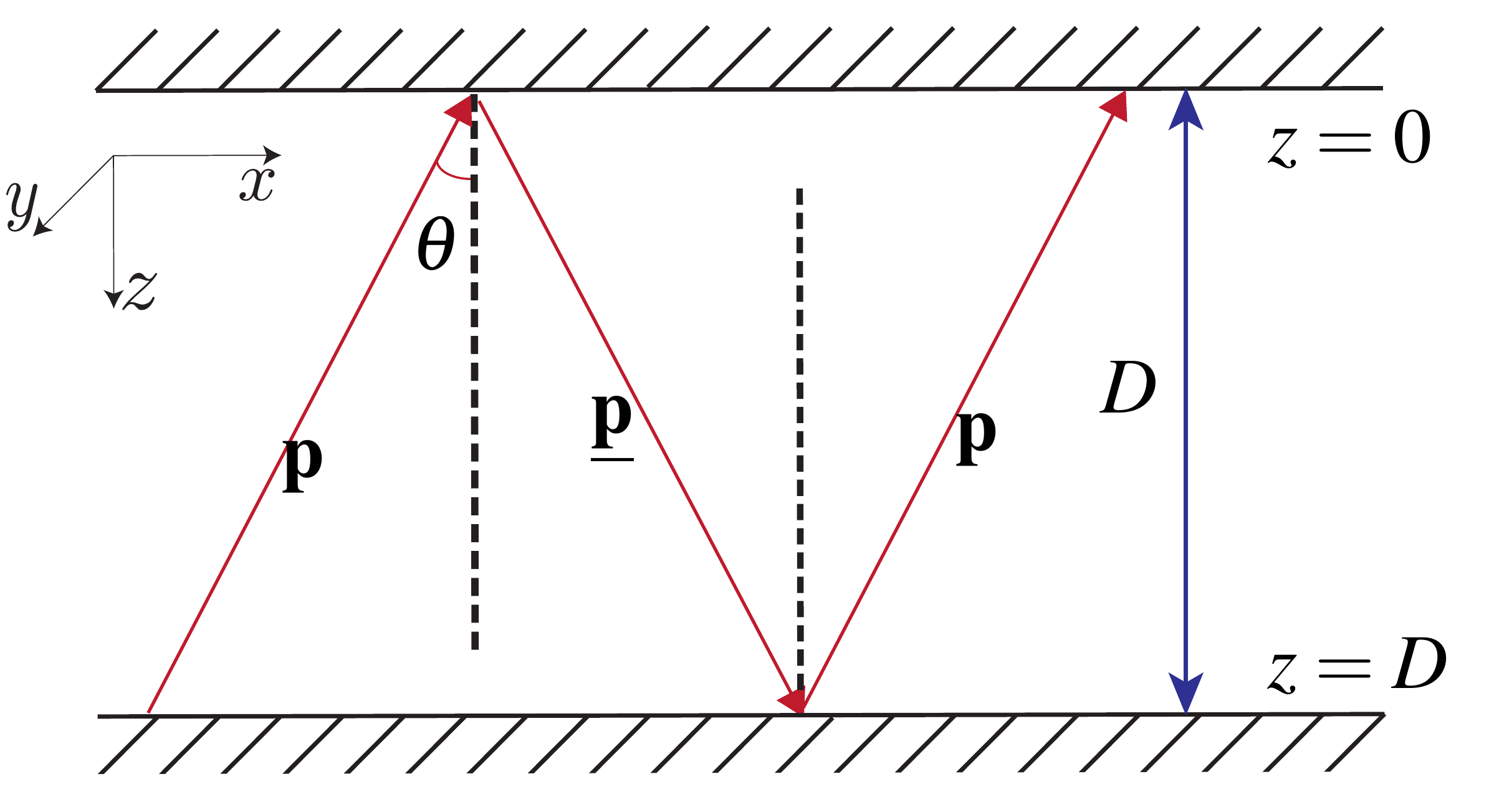}
\caption{Slab geometry of finite width $D$. Trajectories $\vp$ and $\ul{\vp}$ are specularly reflected pairs for which 
the propagators are continuous at both $z = 0$ and $z = D$ surfaces.}
\label{fig-DSR}
\end{figure}

However, for \Heb\ confined in one direction we find that there is no mini-gap in the spectrum (i.e. $\delta=0$) even for
strong confinement.
Our result is presented in Appendix \ref{appendix-equation_blocks} where the propagator, and thus the Fermionic spectrum, is
calculated for the case of confinement by two surfaces separated by distance $D > D_c\approx 9\xi_0$. In the case of strong
confinement we must include reflections, $\vp\rightarrow \ul\vp$ at $z=0$ and $\ul\vp\rightarrow \vp$ at $z=D$ as
shown in Fig. \ref{fig-DSR}.

%------------------------------------ Slab Geometry --------------------------------------------------------------
%------------------------------------------------------------------------------------------------------------------

Double reflections couple the Majorana modes on the two surfaces, modify the spectral weight and profile of the local
density of states, but do not destroy the zero mode or the linear dispersion of the Majorana fermions with
$\vp_{\parallel}$.
This result is shown explicitly in the solution for the quasiclassical propagator in Eqs.
\ref{eq-propagator_in}-\ref{eq-CTin-} for double reflecting trajectories from opposite surfaces. For any finite $D$ the pole
of the propagator is given by the Dirac cone, Eq. \ref{eq-Dirac_Dispersion}, with a zero mode at $\vp_{\parallel}=0$. The
spectral weight is a maximum on both surfaces ($z=0$ and $z=D$), and reduced by a factor of order 
$e^{-D/2\xi_{\Delta}}$ in the center of the channel.
The absence of a mini-gap in the surface spectrum for strong confinement results from the
orthogonality of the Majorana spinors on the opposing surfaces. In particular, the Nambu spinor for the state localized near
$z=0$ describes a right-handed helical spin state (RHSS) in the coordinate system of Fig. \ref{fig-DSR}, 
while the Nambu spinor for the state localized on the opposing
surface ($z=D$) describes a left-handed helical spin state, which is orthogonal to the RHSS. This result holds for strongly
confined B-phase with a self-consistently determined order parameter profile satisfying
$\Delta_{\perp}(0)=\Delta_{\perp}(D)=0$, $\Delta_{\perp}(z)\ne 0$ elsewhere, $\Delta_{\parallel}(z)\ne 0$ and both gap 
functions symmetric about the mid-plane of the channel.\cite{vor03}

The numerical result of Ref. \onlinecite{tsu11} is at odds with our analytic results, which we have also carried out for a
self-consistently determined order parameter profile in a channel confined in one direction. The origin of the discrepancy is
not known at present, but is perhaps related to the multiple reflections from three or four surfaces in a rectangular
channel.
If we assume that non-specular scattering, or confinement in a rectangular channel with four surfaces, generates a mini-gap
as obtained in Ref. \onlinecite{tsu11}, we can examine the effect of the mini-gap on the surface contribution to the mass
current by modifying the surface bound-state dispersion relation, $\eps^{\text{B}}(\vp_{\parallel})
=c|\vp_{\parallel}|\rightarrow \sqrt{c^2\,|\vp_{\parallel}|^2 + \delta^2}$, in the calculation of the surface bound-state
contribution to the current in Eq. \ref{eq-jM_bound-state}.
The resulting temperature dependence of mass current is only weakly modified by the mini-gap (red circles for
$p=34\,\mbox{bar}$), and barely discernible compared to the $T^3$ power law that results from the Dirac cone as shown in 
Fig. \ref{fig-mass_current}.
%
% Experimental observation of the power law correction from the Dirac spectrum in the fully gapped B-phase would provide  
% direct evidence of surface Majorana Fermions.
%-----
% \vspace*{-7mm}
\section*{Acknowledgements}
\vspace*{-5mm}

This research is supported by the National Science Foundation (Grant DMR-1106315).
We thank David Ferguson and Takeshi Mizushima for discussions and critique during the course of this work.
\begin{appendix}
\section{Solution for the Surface Propagator of \Heb}\label{appendix-equation_blocks}

For the B-phase order parameter in Eq. \ref{eq-d_vector_B-phase} the Nambu matrix form of Eilenberger's equation can be
transformed into three-component, first-order matrix differential equations,
\be\label{eq-Vector_Eilenberger}
\frac{\hbar}{2}\vv_{\vp}\cdot\grad\ket{g_{\mathsf{L,T}}} = \widehat{M}_{\mathsf{L,T}}\ket{g_{\mathsf{L,T}}}
\,,
\ee
where the column vectors
\be
\ket{g_{\mathsf{L}}} = \begin{pmatrix} \g^{-}\\ \f_{\mathsf{L}}^{+}\\ \f_{\mathsf{L}}^{-} \end{pmatrix}
\,,\quad
\ket{g_{\mathsf{T}}} = \begin{pmatrix} \g_{\mathsf{T}}^{-}\\ \f_{\mathsf{T}}^{+}\\ \f_{\mathsf{T}}^{-}\end{pmatrix}
\,,
\ee
are defined in terms of the scalar, longitudinal and transverse components of the quasiparticle and Cooper pair propagators
defined in Eq. \ref{eq-quasiclassical_propagator}. The longitudinal and transverse components of a vector in spin space are
defined with respect to the $\vec{d}$-vector for the anisotropic B-phase (Eq. \ref{eq-d_vector_B-phase}), i.e.
$A_{\mathsf{L}}=\vec{A}\cdot\vec{d}/|\Delta(\vp)|$ and $\vec{A}_{\mathsf{T}} = \vec{A}-A_{\mathsf{L}}\vec{d}/|\Delta(\vp)|$.
The $\pm$ notation corresponds to sum and difference of the upper (particle) and lower (hole) components of the Nambu
propagator, $A^{\pm}=(A \pm A')/2$.
The matrices that couple the propagators in the Matsubara representation are then given by
\ber
\widehat{M}_{\mathsf{L}}
&=&
\begin{pmatrix}		0		&	0		&	-i|\Delta(\vp)|	\\
					0		&	0		&	-\eps_{n}		\\
			i|\Delta(\vp)|	&-\eps_{n}	&	0
\end{pmatrix}
\,,
\\
\widehat{M}_{\mathsf{T}}
&=&
\begin{pmatrix}		0			&	-|\Delta(\vp)|		&		0		\\
				-|\Delta(\vp)|	&		0				&	-\eps_{n}	\\
				  	0			&	-\eps_{n}			&	0
\end{pmatrix}
\,,
\eer
with eigenvalues: $\mu = 0\,,\pm\lambda$, where $\lambda=\sqrt{\eps_{n}^2 + |\Delta(\vp)|^2}$.
The longitudinal eigenvector for $\mu=0$
\be\label{egienvectorL0}
\ket{0;\vp}_{\mathsf{L}} 
= \frac{1}{\lambda}\begin{pmatrix}	i\eps_{n}	 \cr |\Delta(\vp)| \cr 0 \end{pmatrix}
\,,
\ee
generates the bulk equilibrium propagator,
\be
\whmfG_{\mathsf{L},0}(\vp,\eps_{n}) 
=-\frac{\pi}{\lambda}
  \begin{pmatrix}			i\eps_{n}					&		-i\vec{\sigma}\sigma_y\cdot\vec{d}(\vp) \cr
  			 	-i\sigma_y\vec{\sigma}\cdot\vec{d}(\vp)	&		-i\eps_{n}	
  \end{pmatrix}
\,.
\ee
This solution satisfies Eilenberger's normalization condition in Eq. \ref{eq-normalization_condition}.
The eigenvectors corresponding to $\mu=\pm\lambda$ are 
\be\label{egienvectorL+-}
\ket{\pm;\vp}_{\mathsf{L}} = \frac{1}{\sqrt{2}\lambda}
\begin{pmatrix}
	\mp|\Delta(\vp)|	\cr 
	\mp \eps_{n} 		\cr 
	\lambda 
\end{pmatrix}
\,.
\ee
These generate ``exploding'' ($\sim e^{+2\lambda s/v_f}$) and ``decaying'' ($\sim e^{-2\lambda s/v_f}$)
solutions to Eq. \ref{eq-Vector_Eilenberger} as a function of the coordinate $s$ along the trajectory $\vv_{\vp}$, and thus 
are physically relevant only in the vicinity of a boundary.\cite{thu84}
The Nambu propagators corresponding to eigenvectors $\ket{\pm;\vp}_{\mathsf{L}}$ are
\be\label{eigenvectorL+-}
\hspace*{-3mm}
\whmfG_{\mathsf{L}, \pm} = \frac{-\pi}{\sqrt{2}\lambda}
\negthickspace\negmedspace
\begin{pmatrix}
	\mp i|\Delta(\vp)|  	
	& 
	\negthickspace\negthickspace\negthickspace\negthickspace\negthickspace			   	
	(i\vec{\sigma}\sigma_y)\cdot \hd (\lambda \mp \eps_n)
	\\
	(i\sigma_y\vec{\sigma})\cdot \hd (-\lambda \mp \eps_n) 	
	& 
	\negthickspace\negthickspace\negthickspace\negthickspace\negthickspace			
	\pm i |\Delta(\vp)|
\end{pmatrix}
\,,
\ee
where $\hat{d}(\vp)\equiv\vec{d}/|\Delta(\vp)|$ is the unit vector in spin space defining the quantization axis for Cooper
pairs with spin projection $\hat{d}\cdot\vec{S} = 0$. 
These matrices are non-normalizable and anti-commute with the bulk propagator,
\be\label{Nambu_Algebra}
\left(\whmfG_{\mathsf{L},\pm}\right)^2 = 0
\,\,,\qquad
\commutator{\whmfG_{\mathsf{L},0}}{\whmfG_{\mathsf{L},\pm}}_{+} = 0
\,.
\ee

For the transverse vector components, the eigenvectors are given by
\be\label{egienvectorsT}
\ket{0;\vp}_{\mathsf{T}} 
= \frac{1}{\lambda}\begin{pmatrix}	\eps_{n}	 \cr -|\Delta(\vp)| \cr 	0 \end{pmatrix}
\,,
\quad
\ket{\pm;\vp}_{\mathsf{T}} 
= \frac{1}{\sqrt{2}\lambda}
\begin{pmatrix}
	\pm |\Delta(\vp)| \cr 
	\lambda \cr 
	\pm \eps_{n}
\end{pmatrix}
\,,
\ee
with corresponding Nambu matrices
\be
\hspace*{-3mm}
\whmfG_{\mathsf{T},0} = -\frac{\pi}{\lambda}
	\begin{pmatrix}
		\eps_n(\vec{\sigma} \cdot \hg)  					&	-(i\vec{\sigma}\sigma_y)\cdot \hf |\Delta(\vp)|	\cr
		-(i\sigma_y\vec{\sigma})\cdot \hf |\Delta(\vp)|		&  -\eps_n (\vec{\sigma}^{tr} \cdot \hg)
	\end{pmatrix}
\,,
\ee
\be
\hspace*{-3mm}
\whmfG_{\mathsf{T},\pm} \negthickspace=\negthickspace \frac{-\pi}{\sqrt{2}\lambda}\negthickspace\negmedspace
	\begin{pmatrix}
		\pm\Delta(\vec{\sigma}\cdot \hg) 
		\negthickspace					
		& 
		\negthickspace
		-(i\vec{\sigma}\sigma_y)\cdot\hf(\lambda\mp\eps_n) 
		\cr
		-(i\sigma_y\vec{\sigma})\cdot\hf(\lambda\pm\eps_n) 	
		\negthickspace\negthickspace\negthickspace\negthickspace\negthickspace
		& 
		\negthickspace\negthickspace\negthickspace\negthickspace\negthickspace
		\mp\Delta(\vec{\sigma}^{tr}\cdot\hg)
	\end{pmatrix}
.
\ee
For the transverse components we introduce an orthonormal basis in spin space for each trajectory $\vp$:
$\{\hat{d},\hat{f}, \hat{g}\}$. The two transverse directions will be fixed by boundary conditions. The transverse matrices 
also satisfy Eqs. \ref{Nambu_Algebra}.

Although $\whmfG_{\mathsf{T},0}$ is normalizable it is not realized in bulk \Heb, and does not contribute to general
solution for confined \Heb. 
However, the transverse exploding and decaying solutions are coupled to the longitudinal components by the surface boundary
condition, and thus play an important role in defining the spectral functions for quasiparticles and Cooper pairs near the
boundary of \Heb.

For the slab geometry of width $D$ shown in Fig. \ref{fig-DSR}, and specular reflection on both surfaces, the propagators
defined on trajecotries $\vp$ and $\ul{\vp}=\vp -2\hat{z}(\hat{z}\cdot\vp)$ have the form,
\ber
\label{eq-propagator_in}
&&
\hspace*{-7mm}
\whmfG^{\text{in}}(\vp,\eps_{n}) = \whmfG_{\mathsf{L,0}}(\vp,\eps_{n}) 
\\
&+& \negthickspace\negthickspace
e^{-2\lambda\,z/\hbar v_z} \left [ C_{\mathsf{L, +}}^{\text{in}}(\vp)\,\whmfG_{\mathsf{L,+}}(\vp,\eps_{n})\,
\negthickspace + \negthickspace
C_{\mathsf{T, +}}^{\text{in}}(\vp)\,\whmfG_{\mathsf{T,+}}(\vp,\eps_{n})\,\right ] 
\nonumber\\
&+&\negthickspace\negthickspace
e^{+2\lambda\,z/\hbar v_z} \left [ C_{\mathsf{L, -}}^{\text{in}}(\vp)\,\whmfG_{\mathsf{L,-}}(\vp,\eps_{n})\,
\negthickspace + \negthickspace
C_{\mathsf{T, -}}^{\text{in}}(\vp)\,\whmfG_{\mathsf{T,-}}(\vp,\eps_{n})\,\right ]
\,,
\nonumber
\eer

\ber
\label{eq-propagator_out}
&&
\hspace{-7mm}
\whmfG^{\text{out}}(\underline{\vp},\eps_{n}) = \whmfG_{\mathsf{L,0}}(\underline{\vp},\eps_{n}) 
\\
&+&\negthickspace\negthickspace
e^{-2\lambda\,z/\hbar\ul{v}_z}
\left [ C_{\mathsf{L, -}}^{\text{out}}(\underline{\vp})\,\whmfG_{\mathsf{L,-}}(\underline{\vp},\eps_{n})\,
\negthickspace + \negthickspace
C_{\mathsf{T, -}}^{\text{out}}(\underline{\vp})\,\whmfG_{\mathsf{T,-}}(\underline{\vp},\eps_{n})\right ]\,
\nonumber\\
&+&\negthickspace\negthickspace
e^{+2\lambda\,z/\hbar\ul{v}_z}
\left [ C_{\mathsf{L, +}}^{\text{out}}(\underline{\vp})\,\whmfG_{\mathsf{L,+}}(\underline{\vp},\eps_{n})\,
\negthickspace + \negthickspace
C_{\mathsf{T, +}}^{\text{out}}(\underline{\vp})\,\whmfG_{\mathsf{T,+}}(\underline{\vp},\eps_{n})\right ]\,,
\nonumber
\eer
where $v_z$ and $\ul{v}_z>0$.
Continuity of the propagators with trajectories $\vp$ and $\ul{\vp}$ at both $z = 0$ and $z = D$ boundaries determines the
coefficients
\ber
\label{eq-CLin+}
&&
\hspace*{-12mm}
C_{\mathsf{L, +}}^{\text{in}} = - C_{\mathsf{L, -}}^{\text{out}}
\\ 
\label{eq-CLin-}
&=&
-\,\frac{\sqrt{2}\,\eps_{n}\,\Delta^2_\perp\cos^2\theta}{|\Delta(\vp)| (\eps_{n}^2 + \Delta_\parallel^2\sin^2\theta)}
\,
\frac{e^{2\lambda\,D/\hbar v_z} - 1}{e^{2\lambda\,D/\hbar v_z} - e^{-2\lambda\,D/\hbar v_z}}
\,,
\nonumber
\eer
\ber
&&
\hspace*{-12mm}
C_{\mathsf{L, -}}^{\text{in}} = - C_{\mathsf{L, +}}^{\text{out}} 
\\
&=&
+\,\frac{\sqrt{2}\,\eps_{n}\,\Delta^2_\perp\cos^2\theta}{|\Delta(\vp)| (\eps_{n}^2 + \Delta_\parallel^2\sin^2\theta)}
\,
\frac{1 - e^{-2\lambda\,D/\hbar v_z}}{e^{2\lambda\,D/\hbar v_z} - e^{-2\lambda\,D/\hbar v_z}}
\,,
\nonumber
\eer
\ber
&&
\label{eq-CTin+}
\hspace{-7mm}
C_{\mathsf{T, +}}^{\text{in}} = - C_{\mathsf{T, -}}^{\text{out}}
\\
&=& 
\frac{\sqrt{2}\sqrt{\eps_{n}^2+|\Delta(\vp)|^2}
     \Delta_\perp\Delta_\parallel\sin\theta\cos\theta}{|\Delta(\vp)|(\eps_{n}^2 + \Delta_\parallel^2\sin^2\theta)}\,
     \frac{e^{2\lambda\,D/\hbar v_z} - 1}{e^{2\lambda\,D/\hbar v_z} - e^{-2\lambda\,D/\hbar v_z}}
\nonumber
\eer
\ber
&&
\label{eq-CTin-}
\hspace*{-7mm}
C_{\mathsf{T, -}}^{\text{in}} = - C_{\mathsf{T, +}}^{\text{out}} 
\\
&=& 
\frac{\sqrt{2}\sqrt{\eps_{n}^2+|\Delta(\vp)|^2}
      \Delta_\perp\Delta_\parallel\sin\theta\cos\theta}{|\Delta(\vp)|(\eps_{n}^2 + \Delta_\parallel^2\sin^2\theta)}\,
      \frac{1 - e^{-2\lambda\,D/\hbar v_z}}{e^{2\lambda\,D/\hbar v_z} - e^{-2\lambda\,D/\hbar v_z}}\,,
\nonumber
\eer
as well as the transverse coordinate axes in spin space: $\hg = \hat{z}\times\hat{d}$ ($\ve_2$ in Fig.
\ref{fig-trajectories}) and $\hf = (\hat{z}\times\hat{d})\times\hat{d}$ for each $\vp$. 

For a surface at $z = 0$ that is well separated from the other surface, i.e. $D \gg \xi_\Delta$, we 
take $D\rightarrow\infty$ in the above solution for slab geometry and obtain the propagator for an isolated 
specular surface with coefficients,
\ber
&&
\hspace{-9mm}
C_{\mathsf{L, +}}^{\text{in}} = - C_{\mathsf{L, -}}^{\text{out}} 
=
-\frac{\sqrt{2}\,\eps_{n}\Delta^2_\perp\cos^2\theta}{|\Delta(\vp)|(\eps_{n}^2 + \Delta_\parallel^2\sin^2\theta)}
\,,
\\
&&
\hspace{-9mm}
C_{\mathsf{T, +}}^{\text{in}} = - C_{\mathsf{T, -}}^{\text{out}} 
\negthickspace = \negthickspace 
\,
\frac{\sqrt{2}\sqrt{\eps_{n}^2+|\Delta(\vp)|^2}
      \Delta_\perp\Delta_\parallel\sin\theta\cos\theta}{|\Delta(\vp)|(\eps_{n}^2 + \Delta_\parallel^2\sin^2\theta)}
\,,
\\
&&
\hspace{-9mm}
C_{\mathsf{L, +}}^{\text{in}} = - C_{\mathsf{L, -}}^{\text{out}} 
=
C_{\mathsf{T, +}}^{\text{in}} = - C_{\mathsf{T, -}}^{\text{out}} =0
\,.
\eer
The vanishing of coefficients, 
$C_{\mathsf{L, +}}^{\text{in}}$, $C_{\mathsf{L, -}}^{\text{out}}$, $C_{\mathsf{T, +}}^{\text{in}}$ and 
$C_{\mathsf{T, -}}^{\text{out}}$ reflects the exclusion of unphysical solutions that would explode into the bulk. 
Analytic continuation ($i\eps_{n}\rightarrow \epsR$) of this solution for an isolated surface gives Eqs. 
\ref{eq-normal_propagator_scalar}-\ref{eq-anomalous_propagator_x} for the components of the retarded propagator.
Note also that for any finite thickness of the channel, the states confined near the two surfaces have a common dispersion 
relation given by the Dirac cone, Eq. \ref{eq-Dirac_Dispersion}.
\smallskip

\section{Sheet spin current}\label{Appendix-Sheet_Spin-Current}

The magnitude of ground state sheet spin current defined by Eq. \ref{eq-sheet_spin-current_energy_integral}
reduces to the integral over the trajectory angles,
\be
I_{\mbox{\footnotesize K}}\left( \frac{\Delta_\parallel}{\Delta_\perp}\right) 
= 
\frac{1}{4}\int_0^{\pi/2}\negthickspace d\theta\sin^2\theta\cos\theta\,
           \tan^{-1}\left(\frac{\Delta_\perp \cos\theta}{\Delta_\parallel\sin\theta}\right)
\,.
\ee
Integration by parts and a change of variables, $u=\cos\theta$ and $x=\Delta_{\parallel}/\Delta_{\perp}$, gives
\ber
I_K(x) 
&=& 
\frac{1}{12x}\,\int_0^1 du\,\,\frac{1-u^2}{x^2 - (x^2 - 1) u^2}
\nonumber\\
&=& \frac{1}{12x(x^2 -1)}\left(1 - \frac{\ln\left(x + \sqrt{x^2 - 1}\right)}{x\sqrt{x^2 - 1}}\right)
\,.
\eer
The two limits of interest are weak confinement ($\Delta_{\perp}\rightarrow\Delta_{\parallel}$), for which 
we evaluate the limit using L'H\^opital's rule,
\be
\lim_{x\rightarrow 1}I_{\mbox{\footnotesize K}}(x) = \frac{1}{18}
\,,
\ee
and strong confinement ($\Delta_{\perp}\ll\Delta_{\parallel}$),
\be
I_{\mbox{\footnotesize K}}(x) \approx \frac{1}{12\,x^3}
\,,\quad x\gg 1
\,.
\ee
Thus, the limiting ground state sheet spin currents are
\be
\mbox{K}(0) = \frac{1}{18}\,N_f v_f^2\hbar^2
\Bigg\{
\begin{array}{ll}
1 &\,,\, \Delta_{\parallel}=\Delta_{\perp}
\cr
\frac{3}{2}\,\left(\Delta_{\perp}/\Delta_{\parallel}\right)^3 &\,,\, \Delta_\perp\ll\Delta_{\parallel}
\end{array}
\,.
\ee
\end{appendix}

%\bibliographystyle{plain}
% \bibliographystyle{apsrev}
%\bibliography{QFS,CM,Books}

%-----
\end{document}